\documentclass[prd,amsmath,amssymb,aps,onecolumn,nofootinbib]{revtex4-1}

\usepackage{subcaption}
\usepackage[dvipsnames]{xcolor}
\usepackage{graphicx}\usepackage{comment}
\usepackage[colorlinks=true,citecolor=blue]{hyperref}
\usepackage{ulem}
\usepackage{soul}
\setstcolor{Green}
\usepackage{cancel}
\usepackage{mathrsfs}
\usepackage{tcolorbox}

\newtcolorbox{questionbox}{
  colback=red!5!white,
  colframe=questionred,
  title=\textbf{Question},
  fonttitle=\sffamily\bfseries
}

\begin{document}

\title{Phenomenology of dynamical trapped regions}

\author{Julio Arrechea}
	\email{julio.arrechea@sissa.it}
	\affiliation{IFPU, Institute for Fundamental Physics of the Universe, via Beirut 2, 34014 Trieste, Italy}
    \affiliation{SISSA, International School for Advanced Studies,
     via Bonomea 265, 34136 Trieste, Italy}
    \affiliation{INFN Sezione di Trieste,
     via Valerio 2, 34127 Trieste, Italy}
\author{Stefano Liberati}
    \email{liberati@sissa.it}
    \affiliation{IFPU, Institute for Fundamental Physics of the Universe, via Beirut 2, 34014 Trieste, Italy}
    \affiliation{SISSA, International School for Advanced Studies,
     via Bonomea 265, 34136 Trieste, Italy}
    \affiliation{INFN Sezione di Trieste,
     via Valerio 2, 34127 Trieste, Italy}   
\author{Hooman Neshat}
    \email{hneshat@sissa.it}
    \affiliation{IFPU, Institute for Fundamental Physics of the Universe, via Beirut 2, 34014 Trieste, Italy}
    \affiliation{SISSA, International School for Advanced Studies,
     via Bonomea 265, 34136 Trieste, Italy}
    \affiliation{INFN Sezione di Trieste,
     via Valerio 2, 34127 Trieste, Italy}
\author{Vania Vellucci}
    \email{vellucci@qtc.sdu.dk}
    \affiliation{Quantum Theory Center ($\hbar$QTC) \& D-IAS, IMADA at Southern Denmark Univ.,\\ Campusvej 55, 5230 Odense M, Denmark.}

\begin{abstract}
Regular black holes (RBHs) and horizonless black hole mimickers (BHMs) are often studied as stationary alternatives to classical black holes, although their interiors may be unstable or undergoing relaxation. We investigate the phenomenological consequences of this evolution by numerically evolving linear scalar perturbations on prescribed, time-dependent Hayward-like geometries. We consider BHM--RBH--BHM histories in which an initially horizonless object temporarily develops a trapped region before returning to a horizonless configuration, through either a single bounce or a sequence of repeated bounces. Using horizon-penetrating Painlev\'e--Gullstrand coordinates, we follow the perturbations through the formation and disappearance of trapping horizons and compute the resulting waveforms and scalar-field energy. For histories with identical initial and final geometries, we find that changing the duration of the intermediate regular black hole phase produces differences in the amplitude and phase of the late-time signal, including its echoes. Longer trapped phases also yield greater energy amplification, consistent with the blueshift of outgoing modes near the inner trapping horizon. These results show that the late-time response of a dynamically relaxing BHM need not be determined by its final configuration alone: it can retain a memory of the dynamical history of its interior.
\end{abstract}

\maketitle


\section{Introduction}
Black holes are traditionally regarded as remaining largely unchanged over long timescales once they have formed. This picture rests on their classical stability in general relativity (GR) and on the slow, adiabatic character of semiclassical Hawking evaporation. However, advances in our understanding of black-hole interiors and their possible instabilities are substantially changing this view.

Under suitable convergence conditions, singularity theorems establish that gravitational collapse leads to a breakdown of classical predictability inside trapping horizons, a conclusion that also extends to gravitational theories beyond GR. At the same time, a substantial body of work across different approaches to quantum gravity suggests that such singular behavior may generically be resolved, giving rise to regular, smooth spacetimes. This possibility can be explored systematically under the assumption that an effective geometric description remains valid in these regimes. Within this framework, two broad families of geometries emerge~\cite{Carballo-Rubioetal2019, Carballo-Rubioetal2019b}: regular black holes (RBHs)~\cite{1968qtr..conf...87B, Dymnikova:1992ux, Hayward2006, Fan:2016hvf}, which retain inner and outer trapping horizons but replace the singularity with a regular core; and horizonless ultra-compact objects, which have no horizons but remain sufficiently compact to mimic black-hole phenomenology. The latter are therefore often called ``black hole mimickers'' (BHMs)~\cite{Visseretal2009, UrbanoVeermae2018, CardosoPani2019}.\footnote{RBHs can be further divided into two broad classes, depending on the presence or absence of wormhole-like throats in their interiors~\cite{SimpsonVisser2018}.}

Singularity resolution also changes the possible causal role of inner horizons. In familiar stationary black-hole geometries, such as Reissner--Nordstr\"om and static RBH spacetimes, the inner horizon coincides with a Cauchy horizon. In a nonsingular dynamical spacetime,
however, an inner trapping horizon need not have this global causal character. Whether a Cauchy horizon ultimately forms depends on the subsequent evolution and global causal structure of the spacetime. In particular, a temporary inner trapping horizon that later disappears need not evolve into a Cauchy horizon.

Stationary inner horizons, which also act as Cauchy horizons, are known to exhibit both classical and semiclassical blueshift instabilities which lead to singular behaviour. When backreaction is taken into account, the classical blueshift instability triggers mass-inflation~\cite{Penrose:1969pc, Poisson:1989zz, Poisson:1990eh, Hamilton:2008zz, Carballo-Rubio:2018pmi, Carballo-Rubio:2021bpr, DiFilippo:2022qkl}. The semiclassical blueshift instability instead reflects the failure of the standard Unruh vacuum to remain regular at the inner horizon~\cite{Flanagan:1997er, Zilbermanetal2022, Balbinot:2023vcm, Carballo-Rubio:2026gwg}, despite its regularity at the outer horizon. The dynamical role of the semiclassical blueshift instability is not yet known in full rigor, although indications of strong singular behaviour have been found~\cite{McMaken:2024fvq}.
A second crucial insight is that, when the stationary description is replaced by a dynamical collapse scenario, these instabilities can instead induce large but regular backreaction effects~\cite{Carballo-Rubio:2026gwg, Arrechea:2026rua}. This suggests that the singular behaviour of inner/Cauchy horizons in stationary geometries should not simply be interpreted as a manifestation of strong cosmic censorship, leaving the stationary exterior intact while the interior terminates at a singularity at the would-be inner horizon. Instead, it may signal that such stationary spacetimes cannot be the endpoints of realistic gravitational collapse. The end-state of semiclassical stellar collapse is still an open question, but it suggests the possibility that the resulting configuration may remain nonsingular but highly dynamical, with its evolution dominated by backreaction, thereby avoiding the formation of a standard GR black hole~\cite{Barenboim:2025ckx, Boyanov:2025otp, Arrechea:2026rua}. 

If this picture is correct, a natural question is what phenomenological signatures arise from these time-dependent backgrounds. Ideally, answering this question requires a unified dynamical treatment of the evolving background and its perturbations. Despite promising developments~\cite{Carballo-RubioNonsingularParadigm2025, Borissova:2026dlz, Borissova:2026wmn}, a complete theory governing the dynamics of such objects is not yet available. RBHs and BHMs are therefore usually studied phenomenologically: one specifies a metric depending on a regularization length scale $L$ and examines how its observational signatures differ from those of its singular counterpart as $L$ is varied~\cite{Franzin:2023slm}. Here, we adopt this phenomenological approach and study linear scalar perturbations on \textit{ad hoc}, time-dependent geometries describing transitions between BHM and RBH phases. We consider two dynamical scenarios of theoretical interest. In the first, a gravitating object near the threshold of horizon formation develops a trapped region that is short-lived from the perspective of an infalling observer before returning to a horizonless configuration. The system thus undergoes a brief ``bounce'', during which a pair of outer and inner trapping horizons forms and subsequently annihilates. In the second scenario, an initially static object undergoes multiple oscillations around the threshold of horizon formation. Depending on their amplitude, these oscillations may or may not produce transient trapped regions, but the system always relaxes to a horizonless configuration in the end.

These scenarios are motivated by recent results showing that trapped regions bounded by outer and inner horizons are unstable in semiclassical gravity~\cite{Barenboim:2025ckx, Boyanov:2025otp}. This instability is dominated by the outward motion of the inner horizon, as predicted using approximations based on dimensional reduction~\cite{Frolov:2016gwl}. As in the classical case, inner-horizon instabilities arise from the strong blueshift of infalling matter and energy fluxes near the horizon, including locally generated fluxes~\cite{Flanagan:1997er}. A crucial difference is that the expectation value of the renormalised stress-energy tensor of quantum fields can violate pointwise energy conditions. The resulting backreaction can evaporate the trapped region on timescales shorter than the Hawking evaporation time, allowing information to escape and potentially producing observable signatures whose properties depend on the timescale and intensity of the process.

A further motivation comes from instabilities under matter accretion, which are sometimes invoked as an argument against the viability of BHMs~\cite{Addazi:2019bjz}. The argument is that, if a BHM has a surface extremely close to its Schwarzschild radius, even a small amount of accreted matter can place that surface inside a trapped region bounded by outer and inner horizons. Understanding the subsequent evolution requires accounting for both classical and semiclassical instabilities. The latter may cause the trapped region to disappear rapidly, opening a relaxation channel towards a horizonless phase~\cite{Barenboim:2025ckx, Boyanov:2025otp, Arrechea:2026rua}. Further work is needed to establish the viability of these speculative scenarios. Here, we draw on these developments to model the phenomenological signatures that may accompany the sudden disappearance of trapped regions using scalar test-fields.

This work also builds on Ref.~\cite{Cardoso:2023guh}, which showed that the inner trapping horizon of a bouncing spacetime can amplify the energy of outgoing radiation. Their numerical analysis used ingoing Eddington--Finkelstein (EF) coordinates and focused on the energy budget, specifically on the relationship between energy amplification, the lifetime of the trapped region, and the surface gravity of the inner horizon. We extend that analysis in three directions. First, we use Painlev\'e--Gullstrand (PG) coordinates~\cite{Painleve:1921, Gullstrand:1922}, which penetrate the horizons and allow us to formulate an evolution problem resembling the standard evolution equations in Schwarzschild coordinates for test fields on static backgrounds. Second, we consider a different class of histories. Instead of interpolating between flat spacetime and a regular black hole, we study BHM--RBH--BHM evolutions, in which an ultra-compact object develops a temporary trapped region before returning to a horizonless state. These histories are relevant to the relaxation scenarios described above and provide a setting in which to explore echoes, a characteristic ringdown signature of horizonless ultra-compact objects. Third, we go beyond energy amplification to compute the full late-time waveform, track the presence and structure of repeated echoes, and examine how the signal depends on the duration of the transient trapped-region phase. Recent work has also explored ringdown signatures of slowly accreting BHMs using similar constructions~\cite{Sharma:2026gvu}.

Our central finding is that the waveform is not determined by the final configuration alone. Two evolutions with identical initial and final ultra-compact geometries, differing only in the lifetime of the intermediate black-hole phase, produce late-time signals with different amplitudes and phases. The ringdown and echo structure therefore retain a memory of the dynamical history of the interior, at least in the idealized, test-field models considered here. This might have direct implications for observations: template banks and echo searches based on stationary horizonless geometries may be systematically mismatched to sources whose interiors are still relaxing. Conversely, the dependence of the signal on the duration of the trapped phase is, in principle, an observable in its own right. The sensitivity anticipated for next-generation ground-based detectors and LISA in the late-inspiral and ringdown regimes motivates a quantitative investigation of these effects.

The paper is organised as follows. In Sec.~\ref{sec:Models}, we introduce the family of dynamical Hayward-like geometries used throughout, obtained by promoting the regularization length scale to a function of the PG time $T$. We then characterise the motion of horizons and light rings for the histories described above. In Sec.~\ref{sec:numerics}, we formulate the scalar-field evolution on these backgrounds, define an energy diagnostic adapted to the PG time slicing, and describe the numerical implementation. We subsequently present results for a static near-extremal black hole mimicker in Sec.~\ref{sec:results}, which serves both as a test of the numerical implementation and as a controlled baseline for studying the echo mechanism, before turning to the fully dynamical BHM--RBH--BHM transitions and the oscillating scenario.

Throughout, we use geometric units $G=c=\hbar=1$ and the metric signature $(-,+,+,+)$. Numerical values of lengths and times are quoted in code units, with the overall physical scale set by the ADM mass $M$. Once $M$ is specified, these quantities can be converted to physical units using the usual geometric-unit conversion. The absolute normalization of the scalar-field energy depends on the chosen initial amplitude, so we focus mainly on relative changes and amplification factors.


\section{Models}
\label{sec:Models}
We are interested in modelling internal relaxation processes of black hole mimickers close to the threshold of horizon formation. The current lack of self-consistent theoretical frameworks with which to test these scenarios motivates the phenomenological approach adopted in this work. Our approach is by no means unique, but is designed to capture the generic features associated with the formation and disappearance of temporary trapped regions. We work directly in PG coordinates, writing
\begin{equation}
\label{eq:PG_metric_models}
    ds^2
    =
    -dT^2
    +
    \left[
        dr+\beta(T,r)\,dT
    \right]^2
    +
    r^2d\Omega^2 ,
\end{equation}
or equivalently,
\begin{equation}
    ds^2
    =
    -F(T,r)\,dT^2
    +
    2\beta(T,r)\,dT\,dr
    +
    dr^2
    +
    r^2d\Omega^2 ,
    \label{eq:PG_metric_expanded_models}
\end{equation}
where \(T\) is the PG time coordinate, \(r\) is the areal radius, and
\begin{equation}
    \beta(T,r)
    =
    \sqrt{1-F(T,r)} .
    \label{eq:beta_models}
\end{equation}
These coordinates are regular across future inner and outer trapping horizons and therefore provide a convenient starting point for describing geometries in which trapped regions form and disappear.

The choice of \(F(T,r)\) adopted in this work is a time-dependent version of the Hayward metric~\cite{Hayward2006},
\begin{equation}
    F(T,r)
    =
    1-\frac{2m(T,r)}{r}.
    \label{eq:Hayward_F_models}
\end{equation}
The corresponding time-dependent Misner--Sharp--Hernandez mass~\cite{MisnerSharp1964, HernandezMisner1966} is
\begin{equation}
    m(T,r)
    =
    \frac{M r^3}{r^3+2M L(T)^2},
    \label{eq:Hayward_mass_models}
\end{equation}
where \(M\) is the ADM mass~\cite{Arnowittetal1960}.\footnote{The Hayward metric is not regular in the strict sense (see the discussion in~\cite{Zhou:2022yio}). While this might result in pathologies when the metric and perturbations are evolved jointly, it does not affect the propagation of test fields over fixed backgrounds.} The time dependence is introduced by promoting the regularization length scale \(L\) to a function of the PG time \(T\).\footnote{We use \(L\) for this length scale in order to avoid confusion with the scalar-field multipole number \(\ell\).} This is at variance with Ref.~\cite{Cardoso:2023guh}, where \(L\) was taken to be constant while \(M\) was promoted to a function of (EF) time, a choice that is better adapted to describe complete formation and evaporation processes.

For fixed \(M\), the critical value of the Hayward length scale is
\begin{equation}
    L_{\rm crit}
    =
    \frac{4M}{3\sqrt{3}} .
    \label{eq:Lcrit_models}
\end{equation}
There are two real positive roots of \(F(T,r)=0\) when \(L(T)<L_{\rm crit}\), one degenerate root when \(L(T)=L_{\rm crit}\), and no roots when \(L(T)>L_{\rm crit}\). Therefore, by varying \(L(T)\), one can interpolate between black hole mimickers and regular black-hole configurations while keeping the ADM mass fixed.

\begin{figure}[h]
    \centering
    \includegraphics[width=0.80\linewidth]{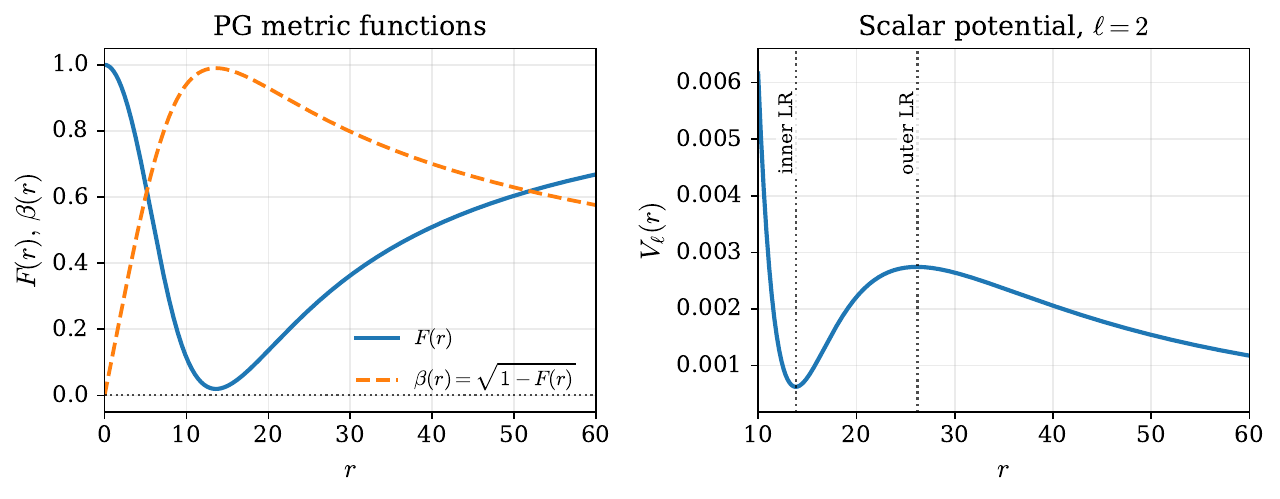}
    \caption{
    Static Hayward black hole mimicker used as the baseline model. The left panel shows the PG metric functions \(F(r)\) and \(\beta(r)=\sqrt{1-F(r)}\), while the right panel shows the scalar effective potential \(V_\ell(r)\) for \(\ell=2\). The parameters are \(M=10\) and \(L=1.03\,L_{\rm crit}\). The vertical dotted lines indicate the inner and outer light-ring positions, which provide a useful guide to the scattering structure of the spacetime.
    }
    \label{fig:model_static_pg_uco}
\end{figure}

Before considering evolving spacetimes, we analyze test propagation on a static horizonless metric as a baseline test of our numerical scheme. The metric used is characterized by
\begin{equation}
    L(T)=L_\textsc{bhm},
    \qquad
    L_\textsc{bhm}>L_{\rm crit}.
\end{equation}
In the simulations below, we consider a near-critical Hayward BHM with
\begin{equation}
    M=10,
    \qquad
    L_\textsc{bhm}=1.03\,L_{\rm crit}.
\end{equation}
The scalar effective potential appearing in the wave equation~\eqref{eq:tortoise_scalar_appendix} reads 
\begin{equation}
    V_\ell(r)
    =
    F(r)
    \left[
        \frac{\ell(\ell+1)}{r^2}
        +
        \frac{F'(r)}{r}
    \right].
    \label{eq:scalar_effective_potential_models}
\end{equation}
This function, together with the metric functions, is shown in Fig.~\ref{fig:model_static_pg_uco}.

The first dynamical model considered in this work describes a single BHM--RBH--BHM transition described by a smooth window function
\begin{equation}
    W(T)
    =
    \frac{1}{2}
    \left[
        \tanh\left(\frac{T-T_{\rm in}}{\Delta T}\right)
        -
        \tanh\left(\frac{T-T_{\rm out}}{\Delta T}\right)
    \right].
    \label{eq:tanh_window_models}
\end{equation}
Here, \(\Delta T\) controls the smoothness of the transitions at \(T_{\rm in}\) and \(T_{\rm out}\). It should not be confused with the duration of the intermediate regular black hole phase, which is instead set by
\begin{equation}
    T_{\rm w}=T_{\rm out}-T_{\rm in}.
\end{equation}

\begin{figure}[h]
    \centering
    \includegraphics[width=0.80\linewidth]{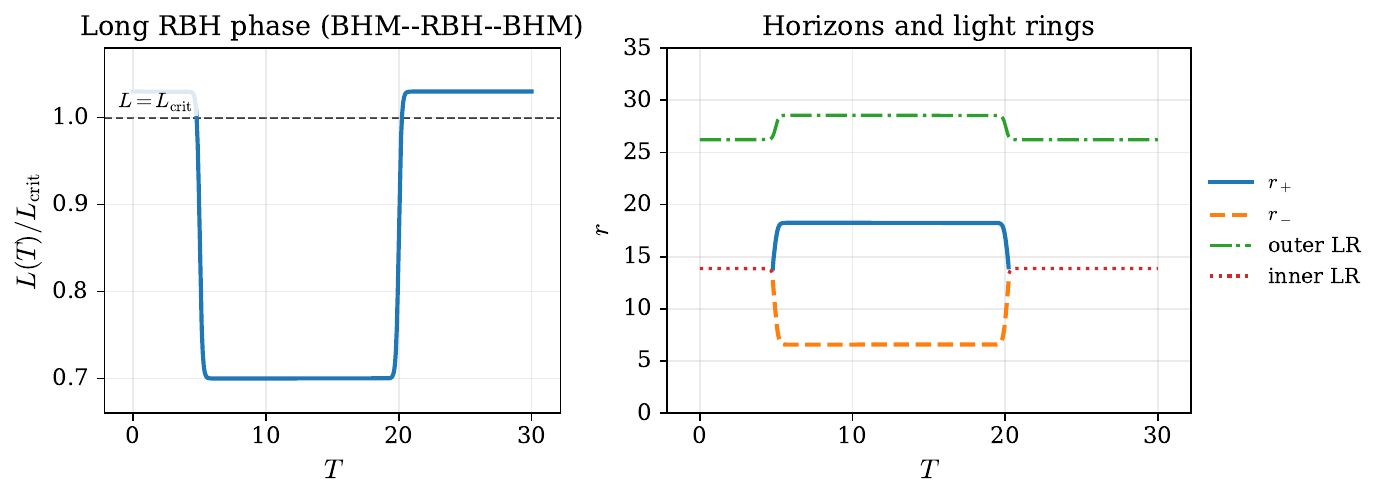}
    \caption{
    Representative BHM--RBH--BHM model in PG time (the difference between ``long'' and ``short'' phases is explained in Sec.~\ref{subsec:dynamical_uco_bh_uco}). The left panel shows the time-dependent regularization scale \(L(T)\), normalized by \(L_{\rm crit}\), for \(M=10\), \(L_\textsc{bhm}=1.03\,L_{\rm crit}\), and \(L_\textsc{rbh}=0.70\,L_{\rm crit}\). The transition is centered around an intermediate regular black hole phase extending approximately over \(T\in[5,20]\), with transition width \(\Delta T=0.20\). The dashed line marks the threshold \(L=L_{\rm crit}\). The right panel shows the corresponding positions of the apparent horizons and instantaneous light rings. Horizons are present only during the intermediate phase in which \(L(T)<L_{\rm crit}\).
    }
    \label{fig:model_pg_uco_rbh_uco}
\end{figure}

Smaller values of \(\Delta T\) correspond to sharper transitions, while larger values produce more gradual ones. We then define
\begin{equation}
    L(T)
    =
    L_\textsc{bhm}
    +
    \left(
        L_\textsc{rbh}-L_\textsc{bhm}
    \right)
    W(T).
    \label{eq:L_single_bounce_models}
\end{equation}
At early times, \(W(T)\simeq0\), so the spacetime is a BHM with \(L(T)\simeq L_\textsc{bhm}\). During the intermediate phase, \(W(T)\simeq1\), so \(L(T)\simeq L_\textsc{rbh}<L_{\rm crit}\), and the geometry contains a trapped region. At late times, \(W(T)\simeq0\) again, and the spacetime returns to the initial BHM configuration. To trace the locations of the outer and inner horizons, \(r_+\) and \(r_-\), we solve
\begin{equation}
    F(T,r)=0 
    \label{eq:horizon_condition_models}
\end{equation}
at fixed \(T\), using the mass function~\eqref{eq:Hayward_mass_models} with the $T$-dependent regularization lengthscale
~\eqref{eq:L_single_bounce_models}. Similarly, we use the condition
\begin{equation}
    \partial_r
    \left(
        \frac{\sqrt{F(T,r)}}{r}
    \right)
    =
    0,
    \label{eq:light_ring_condition_models}
\end{equation}
as an instantaneous diagnostic for the positions of the light rings (in the eikonal approximation~\cite{Carballo-Rubioetal2022b}) in the region where \(F(T,r)>0\). In a strictly static spacetime, this condition gives the circular null geodesics. In the time-dependent case, it should instead be interpreted as a representation of the compactness and scattering structure of the geometry.
An example of this BHM--RBH--BHM transition is shown in Fig.~\ref{fig:model_pg_uco_rbh_uco}. In the numerical analysis below, we use the same values of \(L_\textsc{bhm}\) and \(L_\textsc{rbh}\), but compare different durations of the intermediate trapped-region phase.

\begin{figure}[h]
    \centering
    \includegraphics[width=0.80\linewidth]{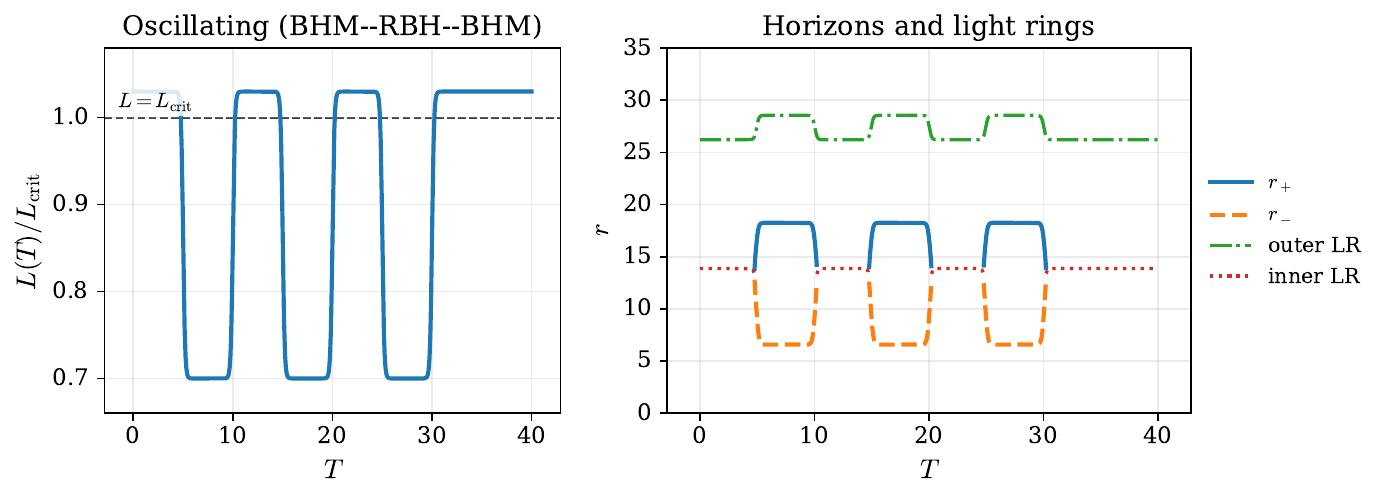}
    \caption{
    Representative oscillating model in PG time. The left panel shows the regularization scale \(L(T)/L_{\rm crit}\) for \(M=10\), \(L_\textsc{bhm}=1.03\,L_{\rm crit}\), and \(L_\textsc{rbh}=0.70\,L_{\rm crit}\), with transition width \(\Delta T=0.20\). The model undergoes three temporary regular black hole phases, approximately over the intervals \(T\in[5,10]\), \(T\in[15,20]\), and \(T\in[25,30]\). The right panel shows the corresponding apparent horizons and instantaneous light rings. Each interval with \(L(T)<L_{\rm crit}\) corresponds to a temporary regular black hole phase bounded by an outer and an inner apparent horizon.
    }
    \label{fig:model_pg_oscillating_case}
\end{figure}

The second dynamical model is the oscillating case, in which the geometry undergoes repeated passages through the threshold of horizon formation. This is implemented by summing several non-overlapping windows,
\begin{equation}
    L(T)
    =
    L_\textsc{bhm}
    +
    \left(
        L_\textsc{rbh}-L_\textsc{bhm}
    \right)
    \sum_{j=1}^{N_{\rm b}}
    W_j(T),
    \label{eq:L_oscillating_models}
\end{equation}
where
\begin{equation}
    W_j(T)
    =
    \frac{1}{2}
    \left[
        \tanh\left(\frac{T-T^{(j)}_{\rm in}}{\Delta T_j}\right)
        -
        \tanh\left(\frac{T-T^{(j)}_{\rm out}}{\Delta T_j}\right)
    \right].
    \label{eq:tanh_window_j_models}
\end{equation}
Each window represents one temporary BHM--RBH--BHM transition. This provides a simple phenomenological model of an object oscillating around its extremal (critical) configuration, repeatedly developing and losing a trapped region before returning to a horizonless state. One of the oscillating models used in the numerical analysis is illustrated in Fig.~\ref{fig:model_pg_oscillating_case}.

The models defined above are deliberately minimal. Keeping \(M\) fixed while varying \(L(T)\) allows us to change the presence or absence of trapped regions without changing the ADM mass between the initial and final configurations, thus allowing us to simulate changes in the interior properties of regions with strong gravitational effects. However, it is very important to note that our dynamical models account for a rapid formation and disappearance of the trapped region, a feature that could be inconsistent with a matter sector that obeys causality. Hence, our construction reflects only qualitatively the dynamical solutions of modified-gravity theories incorporating regular black holes and semiclassical effects (e.g.,~\cite{Taves:2014laa, Bueno:2024dgm, Barenboim:2025ckx, Borissova:2026wmn} and references therein). Having specified the background geometries and their time
dependence, we now turn to the evolution of scalar perturbations on these spacetimes and to the numerical framework used to study them.


\section{Numerical framework}
\label{sec:numerics}

In this section, we describe the numerical framework used to evolve linear scalar perturbations on the time-dependent compact-object geometries introduced in Sec.~\ref{sec:Models}. As mentioned above, we adapt the framework of Refs.~\cite{Cardoso:2023guh, Sharma:2026gvu}, which simulates field propagation in Eddington--Finkelstein coordinates, to Painlev\'e--Gullstrand coordinates. We motivate this choice in what follows.


\subsection{Motivation for Painlev\'e--Gullstrand coordinates}

The background geometries used in the numerical evolution were introduced in Sec.~\ref{sec:Models}. The PG coordinates parametrize time $T$ as measured by clocks associated with freely falling timelike observers, are regular across horizons, and keep $r$ as the physical areal radius, thus making them suitable for evolving perturbations through both horizonless and horizon-containing phases. They also allow us to define a natural energy diagnostic associated with the PG time evolution, which will be used below.

Readers might be familiar with the usual echo description for static stars, which makes use of the tortoise coordinate $r_*$ and the corresponding static time $t$. In those coordinates, the scalar equation takes the form of a one-dimensional scattering problem, and the echo structure can be clearly identified. However, this formulation is not horizon-penetrating. Therefore, it cannot be used as a single coordinate system for the full BHM--RBH--BHM evolution, in which horizons may form and later disappear.

The Painlev\'e--Gullstrand coordinates provide a horizon-penetrating slicing while keeping the evolution in Cauchy form, meaning that initial data can be prescribed on spacelike hypersurfaces. In this way, we can compare different types of initial wave packets, such as ingoing, outgoing, or time-symmetric data. By contrast, in ingoing Eddington--Finkelstein coordinates, which are another horizon-penetrating coordinate system, the initial profile is specified on a constant-$v$ null surface. Although this coordinate system is regular at future horizons and is well suited for describing black-hole formation, it provides less freedom in the choice of initial data. Finally, PG coordinates can be easily generalized to describe black-to-white hole transitions~\cite{Gaur:2023ved}, allowing future exploration of echo signals associated with such processes.

\subsection{Scalar-field equation in PG form}
\label{subsec:scalar_wave_pg}

We consider a test massless scalar field satisfying
\begin{equation}
    \Box \Phi = 0 .
    \label{eq:KG_full}
\end{equation}
Using the spherical symmetry of the background, we decompose the scalar field into spherical harmonics,
\begin{equation}
    \Phi(T,r,\theta,\varphi)
    =
    \frac{\phi(T,r)}{r}
    Y_{\ell m}(\theta,\varphi).
    \label{eq:scalar_decomposition}
\end{equation}
The evolved variable is therefore the reduced radial field
\(\phi(T,r)\). We introduce
\begin{equation}
    \Pi(T,r)
    \equiv
    \partial_T\phi(T,r),
    \label{eq:Pi_def}
\end{equation}
so that the scalar wave equation can be written as a system that is first order in time.

Substituting the PG metric~\eqref{eq:PG_metric_models} and the mode decomposition~\eqref{eq:scalar_decomposition} into
Eq.~\eqref{eq:KG_full}, and using
\begin{equation}
    \Delta_{S^2}Y_{\ell m}
    =
    -\ell(\ell+1)Y_{\ell m},
\end{equation}
where \(\Delta_{S^2}\) denotes the angular part of the Laplacian on the unit two-sphere, gives
\begin{align}
    \partial_T\Pi
    =
    &
    \beta\,\partial_r\Pi
    +
    \partial_r(\beta\Pi)
    +
    F\,\partial_r^2\phi
    +
    \left(
        \beta_T+F_r
    \right)
    \left(
        \partial_r\phi-\frac{\phi}{r}
    \right)
    -
    \frac{\ell(\ell+1)}{r^2}\phi .
    \label{eq:PG_wave_equation_split}
\end{align}
Together with Eq.~\eqref{eq:Pi_def}, this constitutes the evolution system used in the simulations.

All radial derivatives in Eq.~\eqref{eq:PG_wave_equation_split} are
evaluated using second-order finite differences. Evolution between constant-\(T\) slices is performed with a fourth-order Runge--Kutta method, with the time step chosen according to the characteristic propagation speeds of the scalar field. For dynamical backgrounds, \(\beta_T\) is computed from the background geometry, while it vanishes in the static case. The main numerical difficulty arises during the regular-black-hole phase. The strong blueshift near the inner horizon produces increasingly sharp radial structures in the scalar field, which require high spatial resolution in the compact region. We therefore use a radial grid that is much finer close to the object and gradually becomes coarser towards the outer boundary. This allows us to resolve the dynamics near the inner horizon without making the large exterior domain unnecessarily expensive to evolve. If this region is not sufficiently resolved, small-scale numerical oscillations may appear in the field profile; these are reduced as the resolution is increased.


\subsection{Energy diagnostic}
\label{subsec:energy_pg}

The energy diagnostic used in the simulations is obtained from the stress-energy tensor of the massless scalar field,
\begin{equation}
    \mathcal{T}_{\mu\nu}
    =
    \partial_\mu\Phi\,\partial_\nu\Phi
    -
    \frac{1}{2}g_{\mu\nu}
    g^{\alpha\beta}\partial_\alpha\Phi\,\partial_\beta\Phi .
    \label{eq:scalar_stress_tensor}
\end{equation}
Given the PG time coordinate $T$, a natural energy current is associated with the vector field
\begin{equation}
    \xi^\mu=(\partial_T)^\mu .
\end{equation}
In a static geometry, this vector is Killing, and the corresponding energy is conserved up to boundary fluxes and numerical error. In a dynamical
geometry, $\partial_T$ is not a Killing vector, so the same quantity is
not conserved; nevertheless, it remains a useful diagnostic for measuring
the energy stored in the scalar perturbation on a $T=\mathrm{constant}$
slice.

The energy on a constant-$T$ hypersurface is obtained by integrating the current associated with $\xi^\mu$. Equivalently, in coordinate form,
it can be written as
\begin{equation}
    E_{\partial_T}(T)
    =
    -\int_{T=\mathrm{const.}} d^3x\,
    \sqrt{-g}\,
    \mathcal{T}^{T}{}_{T}.
    \label{eq:energy_from_TT}
\end{equation}

The overall sign follows from contracting the energy current with the future-directed unit normal to the constant-$T$ hypersurfaces. In a static horizonless region, where $\partial_T$ is timelike, this quantity coincides with the conserved energy associated with time translations and is positive definite. During a trapped-region phase, however, $\partial_T$ becomes spacelike wherever $F<0$, and the corresponding energy density is no longer positive definite. We therefore use $E_{\partial_T}$ as an energy diagnostic throughout the evolution, while its interpretation as a positive conserved energy applies directly only in the static horizonless phases. For the PG metric~\eqref{eq:PG_metric_expanded_models}, we have
\begin{equation}
    \sqrt{-g}
    =
    r^2\sin\theta ,
\end{equation}
and the inverse metric components in the $(T,r)$ sector are
\begin{equation}
    g^{TT}=-1,
    \qquad
    g^{Tr}=\beta,
    \qquad
    g^{rr}=F .
\end{equation}
A direct substitution into Eq.~\eqref{eq:scalar_stress_tensor} gives the
useful cancellation of the shift terms,
\begin{equation}
    -\mathcal{T}^{T}{}_{T}
    =
    \frac{1}{2}
    \left[
        (\partial_T\Phi)^2
        +
        F(\partial_r\Phi)^2
        +
        \frac{1}{r^2}
        |\nabla_{S^2}\Phi|^2
    \right].
    \label{eq:minus_TT_scalar}
\end{equation}
Thus, the energy density contains no explicit cross term proportional to $\partial_T\Phi\,\partial_r\Phi$, even though the PG metric has a nonzero shift.

Using the spherical-harmonic decomposition
\begin{equation}
    \Phi(T,r,\theta,\varphi)
    =
    \frac{\phi(T,r)}{r}Y_{\ell m}(\theta,\varphi),
    \label{eq:harmonic_decomposition}
\end{equation}
together with
\begin{equation}
    \int d\Omega\, |Y_{\ell m}|^2=1,
    \qquad
    \int d\Omega\, |\nabla_{S^2}Y_{\ell m}|^2
    =
    \ell(\ell+1),
\end{equation}
and using $\Pi=\partial_T\phi$, we obtain
\begin{equation}
    \partial_T\Phi
    =
    \frac{\Pi}{r}Y_{\ell m},
    \qquad
    \partial_r\Phi
    =
    \frac{1}{r}
    \left(
        \partial_r\phi-\frac{\phi}{r}
    \right)Y_{\ell m}.
\end{equation}
After integrating over the angles, the energy of a single $(\ell,m)$
mode becomes
\begin{equation}
    E_{\partial_T}(T)
    =
    \frac{1}{2}
    \int_0^\infty dr
    \left[
        \Pi^2
        +
        F
        \left(
            \partial_r\phi-\frac{\phi}{r}
        \right)^2
        +
        \frac{\ell(\ell+1)}{r^2}
        \phi^2
    \right].
    \label{eq:PG_energy}
\end{equation}

We use this quantity as a conservation check in static horizonless evolutions and as a measure of the net energy amplification between the initial and final static phases of the dynamical geometries.


\subsection{Boundary conditions and regularity at the center}

The radial evolution domain is taken to be
\begin{equation}
    r\in[0,r_{\rm max}],
\end{equation}
where $r_{\rm max}$ is the outermost boundary of a fiducial numerical grid intended to represent radial infinity. At the center, regularity of the full scalar field fixes the allowed behavior of the reduced field $\phi$. Since
\begin{equation}
    \Phi(T,r,\theta,\varphi)
    =
    \frac{\phi(T,r)}{r}Y_{\ell m}(\theta,\varphi),
\end{equation}
regularity requires the reduced field to vanish at least linearly at the center.
Using the near-origin form of Eq.~\eqref{eq:PG_wave_equation_split},
which contains the angular term proportional to $\ell(\ell+1)/r^2$,
then selects the regular behaviour
\begin{equation}
    \phi(T,r)\sim r^{\ell+1}
    \qquad
    \text{as}
    \qquad
    r\to0 .
\end{equation}
In particular, the reduced field and its momentum must vanish at the origin for $\ell>0$,
\begin{equation}
    \phi(T,0)=0,
    \qquad
    \Pi(T,0)=0 .
\end{equation}
This condition implements the regular reflecting center of the geometry. Numerically, derivatives near the origin are evaluated using the corresponding parity of the regular solution, so that no singular behaviour is introduced by the explicit factors of $1/r$ appearing in the equations.

At the outer boundary $r=r_{\rm max}$, we impose an outgoing condition, allowing radiation to leave the computational domain without reflection. In PG coordinates, outgoing radial characteristics have coordinate speed
\begin{equation}
    \frac{dr}{dT}
    =
    (1-\beta),
\end{equation}
and the boundary condition is implemented in the form
\begin{equation}
    \Pi
    \simeq
    -(1-\beta)\,\partial_r\phi
    \qquad
    \text{at}
    \qquad
    r=r_{\rm max}.
\end{equation}
This is the PG analogue of a Sommerfeld outgoing wave condition~\cite{Sommerfeld1912, BaylissTurkel1980}. It therefore enforces outgoing propagation at \(r=r_{\rm max}\), with no incoming characteristic mode prescribed there. 

These boundary conditions together ensure that the inner boundary represents a smooth, regular center, while the outer boundary approximates an open system in which radiation can propagate outward.


\subsection{Initial data}

The scalar perturbation is initialized as a localized wave packet. We use a regularized Gaussian profile of the form
\begin{equation}
    \phi(0,r)
    =
    A
    \left(
        \frac{r}{\sqrt{r^2+r_{\rm reg}^2}}
    \right)^{\ell+1}
    \exp\left[
        -\frac{(r-r_0)^2}{\sigma^2}
    \right],
    \label{eq:initial_gaussian}
\end{equation}
where $A$ is the amplitude, $r_0$ is the initial location of the packet, $\sigma$ is its width, and $r_{\rm reg}$ is a small regulator used only to make the regular behavior at the origin explicit. The initial momentum
\begin{equation}
    \Pi(0,r)=\partial_T\phi(0,r)
\end{equation}
determines the initial direction of propagation. Initial data containing both ingoing and outgoing components are obtained by setting
\begin{equation}
    \Pi(0,r)=0 .
\end{equation}
Although this choice does not correspond to equal ingoing and outgoing amplitudes in PG coordinates, it excites both propagation sectors. 

To prepare a mostly ingoing packet, we instead choose $\Pi(0,r)$ such that the initial profile follows an ingoing radial characteristic. In PG coordinates, this is implemented as
\begin{equation}
    \Pi(0,r)
    \simeq
    (1+\beta)\,\partial_r\phi(0,r).
    \label{eq:initial_ingoing}
\end{equation}
Similarly, a mostly outgoing packet is obtained with
\begin{equation}
    \Pi(0,r)
    \simeq
    -(1-\beta)\,\partial_r\phi(0,r).
    \label{eq:initial_outgoing}
\end{equation}

In the simulations below, the initial packet is placed well inside the numerical domain and sufficiently far from the outer boundary. This is an operational simplification that allows us to easily follow the interaction of the field with the inner horizons. With the background models, evolution equations, numerical scheme, and initial data now specified, we turn to the
resulting scalar-field dynamics.


\section{Results}
\label{sec:results}
\subsection{Static black hole mimicker}
\label{subsec:static_uco}

We first consider a static black hole mimicker close to the threshold of horizon formation. This provides a useful baseline before studying fully dynamical BHM--RBH--BHM transitions. In this case, the geometry is time-independent, so the scalar-field energy is expected to be conserved up to boundary fluxes and numerical error, while the waveform can be used to study the standard echo mechanism in a controlled setting.

The initial wave packet is placed within the compact interior region, inside the outer maximum of the effective potential. Since the initial data are chosen with $\Pi(0,r)=0$, the perturbation contains both ingoing and outgoing components. The outgoing part propagates towards the outer potential barrier, where it is partially transmitted to the exterior and partially reflected back towards the interior. The ingoing part instead propagates towards the regular center, where it is reflected and subsequently returns to the potential barrier. Repeated propagation between the center and the outer barrier then produces a sequence of delayed pulses in the observed waveform. The initial position of the wave packet affects the overall structure of the observed field amplitude, including the timing of the different features in the waveform. Nevertheless, for the purposes of our echo analysis, this dependence is not particularly important, since the qualitative echo mechanism remains unchanged. We therefore place the initial wave packet well within the compact interior region, inside the outer maximum of the effective potential.

We compare the time-domain evolution in three coordinate systems: the standard tortoise-coordinate formulation $(t,r_*)$, the PG formulation $(T,r)$, and the ingoing Eddington--Finkelstein formulation $(v,r)$. The first is the usual coordinate system used for static echo calculations, while the latter two are horizon-penetrating and therefore more closely connected to the dynamical evolutions discussed later. The relevant equations for the Eddington--Finkelstein and tortoise-coordinate formulations are displayed in Appendices~\ref{app:EF_scalar} and~\ref{app:tortoise_scalar}, respectively.

\begin{figure}[h]
    \centering
    \begin{subfigure}{0.32\textwidth}
        \centering
        \includegraphics[width=\textwidth]{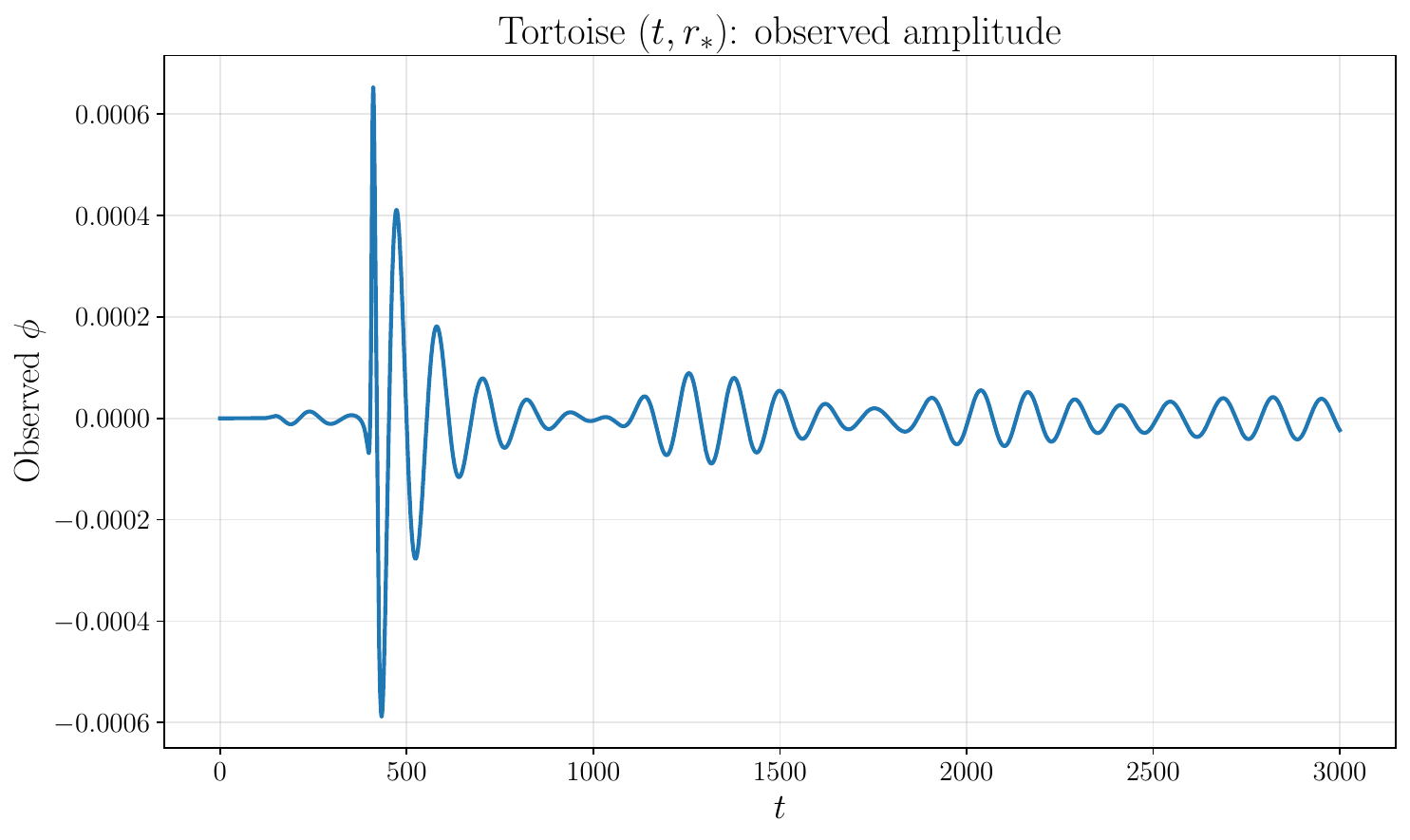}
        \caption{$(t,r_*)$}
    \end{subfigure}
    \hfill
    \begin{subfigure}{0.32\textwidth}
        \centering
        \includegraphics[width=\textwidth]{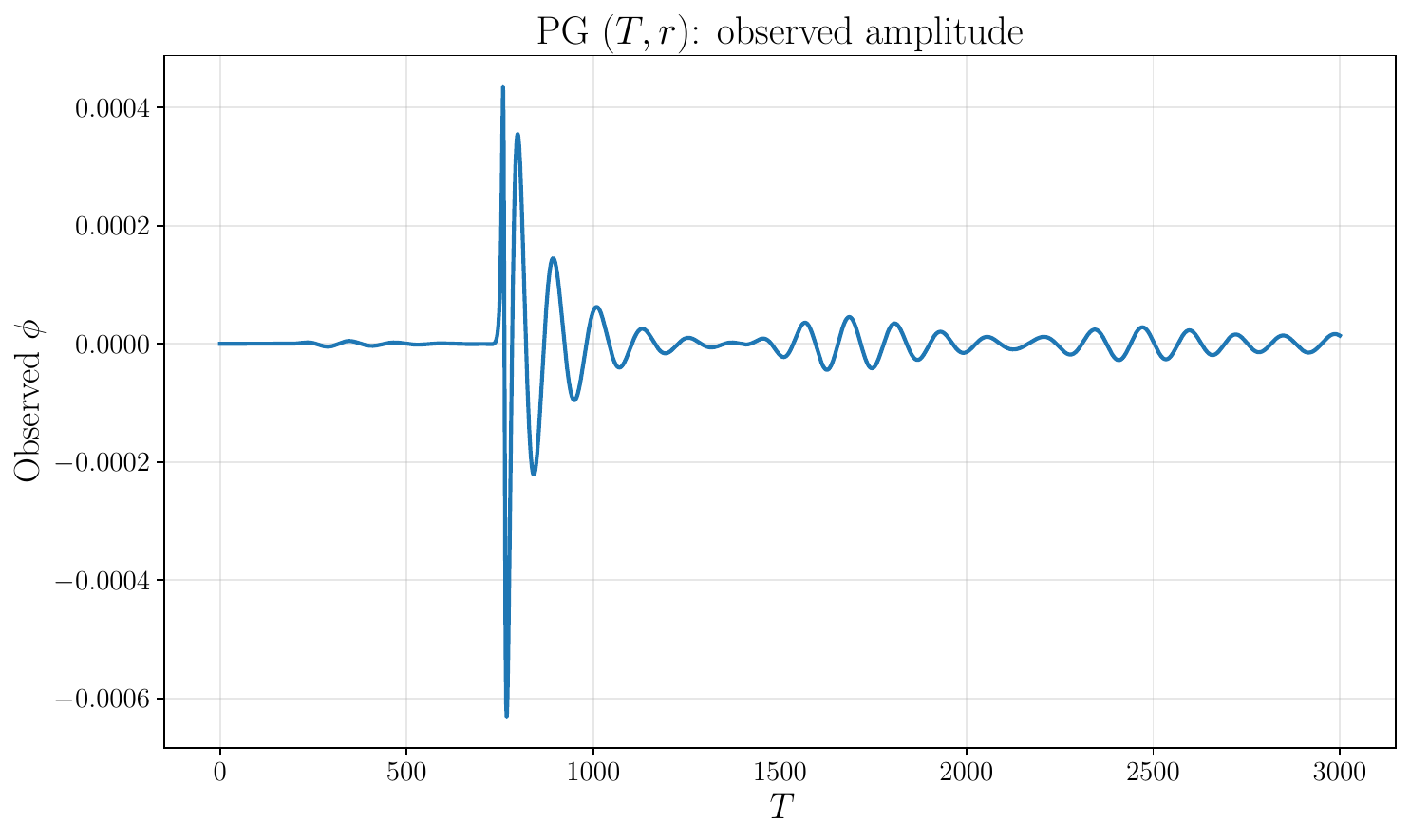}
        \caption{PG $(T,r)$}
    \end{subfigure}
    \hfill
    \begin{subfigure}{0.32\textwidth}
        \centering
        \includegraphics[width=\textwidth]{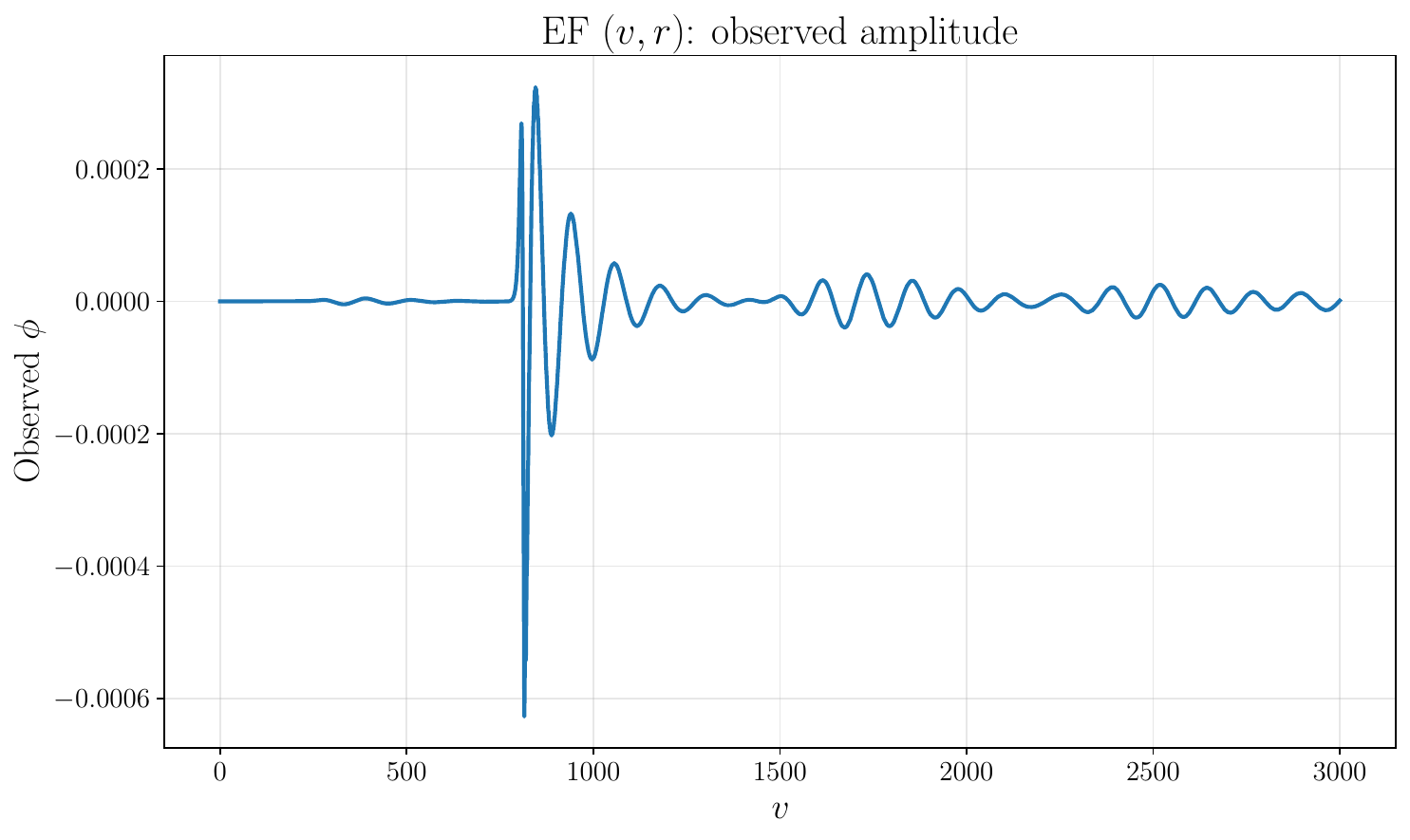}
        \caption{EF $(v,r)$}
    \end{subfigure}
    \caption{
     Observed scalar amplitude for a static Hayward BHM close to extremality, evolved in three different coordinate systems. The model parameters are $\ell=2$, $M=10$, and $L=1.03\,L_{\rm crit}$. The initial perturbation is a regular Gaussian packet with amplitude $A=10^{-3}$, centered at $r_0=10$, and width $\sigma=5$, and the waveform is observed at $r_{\rm obs}=90$. The echo train is visible in all three formulations.
    }
    \label{fig:static_uco_amplitudes}
\end{figure}

The field amplitudes measured by a static observer at a distant radius are shown in Fig.~\ref{fig:static_uco_amplitudes}.  In the $(t,r_*)$ evolution, the echo structure is very clear: after the initial burst, the signal displays a sequence of well-separated delayed pulses. The same qualitative structure is also visible in the PG evolution, showing that the PG formulation is able to capture the echo signal of a static horizonless BHM. This signal is also captured by the $(v,r)$ evolution. Echoes are a physical consequence of repeated scattering between the regular center and the outer potential barrier, and are observed in all three coordinate systems. Compared to the PG formulation, the main limitation of the EF one lies in the choice of initial data. Since the initial profile is specified on a constant-$v$ null surface, one does not have the same freedom to prescribe arbitrary Cauchy data, such as purely ingoing, outgoing, or a mix of the two. In practice, the characteristic initial data are naturally adapted to outgoing radiation.

One can see from the plots that, before the main burst, the waveform exhibits a small oscillatory precursor. This feature originates from the finite radial width of the initial wave packet. The outer part of the initial profile is located outside the region of strongest gravitational delay and therefore reaches the observer earlier than the bulk of the perturbation. The majority of the wave then propagates through the compact region and gives rise to the much larger primary burst. The subsequent, individually resolvable echo patterns correspond instead to radiation that has undergone additional propagation inside the effective cavity and constitutes the echo signal.


\subsection{Dynamical BHM--RBH--BHM}
\label{subsec:dynamical_uco_bh_uco}

We now turn to time-dependent geometries in which an initially horizonless black hole mimicker enters a temporary regular black hole phase and subsequently returns to a horizonless final state, as discussed in Sec.~\ref{sec:Models}. The initial and final configurations are chosen to be the same BHM, while the duration of the intermediate regular black hole phase is varied.

Both models have $\ell=2$, $M=10$, $L_\textsc{bhm}=1.03\,L_{\rm crit}$, and $L_\textsc{rbh}=0.70\,L_{\rm crit}$. The ``short'' model remains in the black-hole phase approximately during the interval $T\in[5,10]$, while the ``long'' model remains in the black-hole phase approximately during $T\in[5,20]$. The transition width $\Delta T$ is kept fixed in both cases. We choose initial data containing both ingoing and outgoing modes, modeled through a Gaussian centered at $r_{0}=10$. We have selected the parameters of the models in a way that better highlights the features of the resulting waveforms.

\begin{figure}[h]
    \centering
    \includegraphics[width=0.70\linewidth]{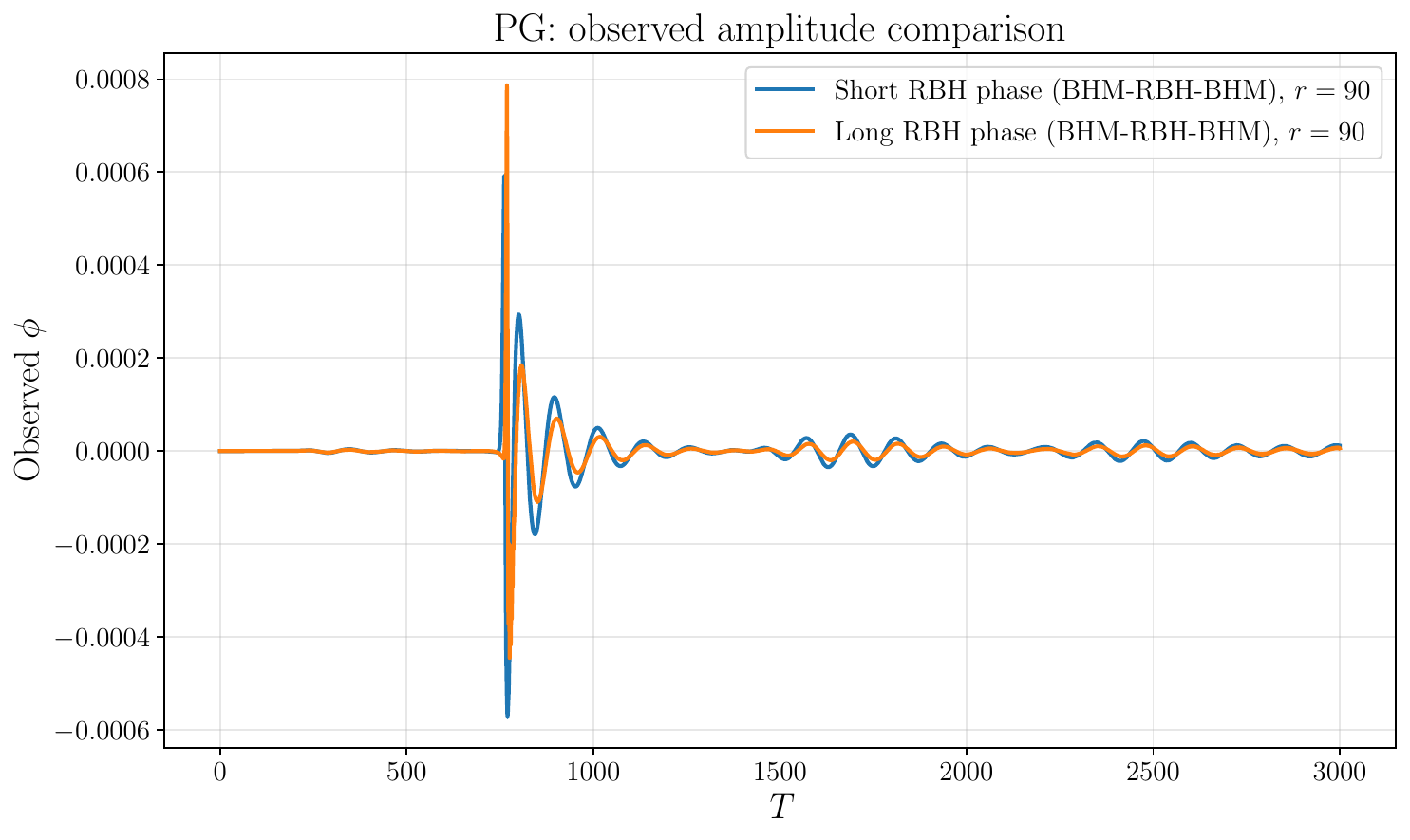}
    \caption{
    Observed scalar amplitude in PG coordinates for two BHM--RBH--BHM evolutions measured at $r=90$. Both models have $\ell=2$, $M=10$, $L_\textsc{bhm}=1.03\,L_{\rm crit}$, and $L_\textsc{rbh}=0.70\,L_{\rm crit}$. The two cases differ only in the duration of the intermediate regular black hole phase: approximately $T\in[5,10]$ for the short case and $T\in[5,20]$ for the long case.
    }
    \label{fig:PG_dynamic_waveform_comparison}
\end{figure}

Figure~\ref{fig:PG_dynamic_waveform_comparison} shows the scalar waveform observed at a large radius ($r_{\textrm{obs}}=90$) in PG coordinates. The two measured signals have a similar qualitative structure, as expected from the fact that the initial and final compact objects are the same. However, their amplitudes and phases differ after the interaction with the time-dependent region. The physical interpretation is that, during the temporary black-hole phase, part of the scalar perturbation propagates through a geometry containing trapped regions and an inner-horizon structure. Increasing the duration of this phase gives the perturbation more time to interact with the trapped region before the spacetime returns to the final BHM configuration. Consequently, the late-time waveform is not determined solely by the final compact-object configuration, but can retain information about the system's preceding dynamical history.

\begin{figure}[h]
    \centering
    \includegraphics[width=0.6\linewidth]{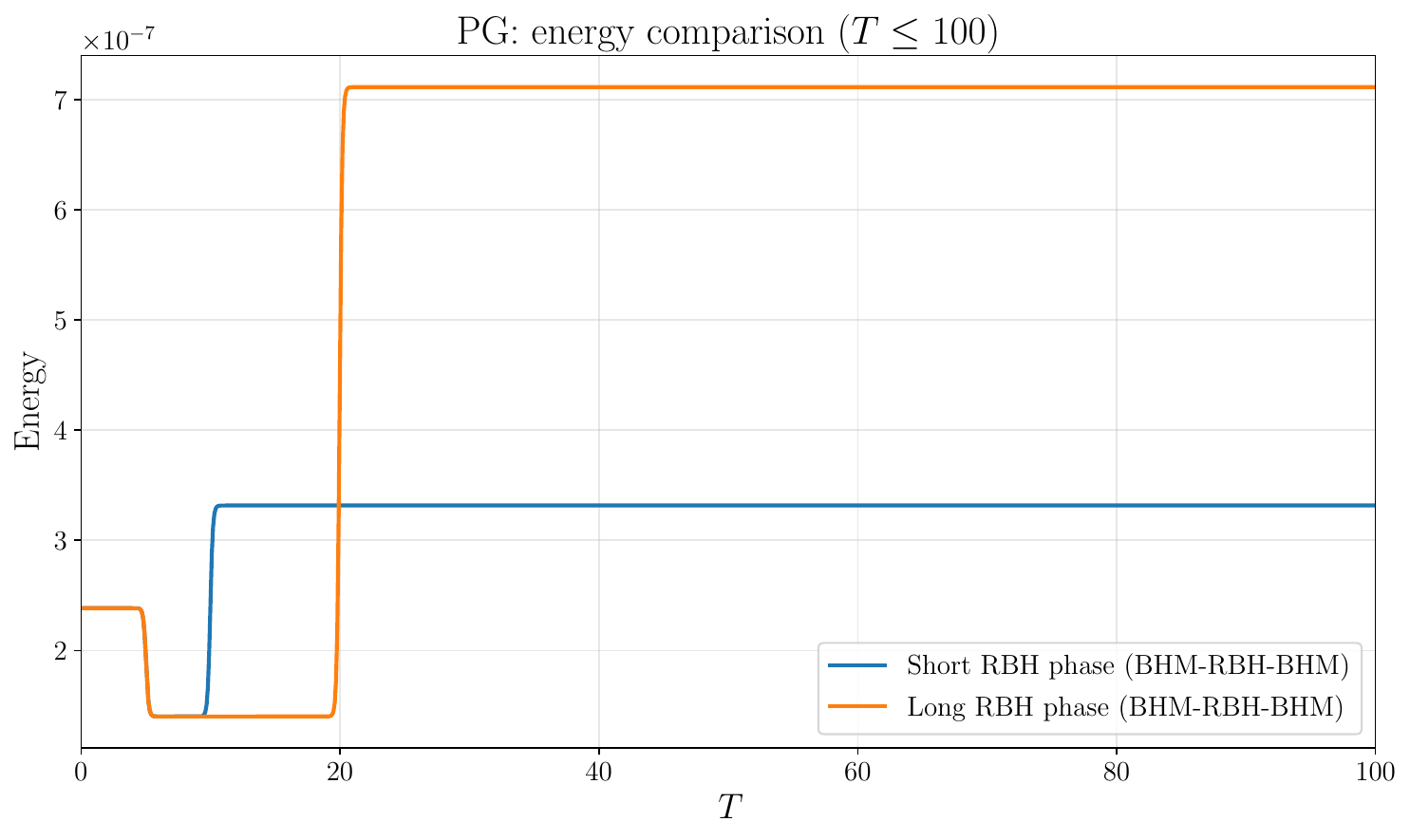}
    \caption{Field energy comparison between the ``short RBH'' and ``long RBH'' models for the first \(100\) units of PG code time. The intermediate values should not be interpreted as a conserved physical energy during the time-dependent phases; the meaningful comparison is between the initial and final static regions.}
    \label{fig:PG_energy_comparison}
\end{figure}

In addition to the late-time differences shown between the two models, one naturally expects a difference in the energy associated with each scalar field~\eqref{eq:PG_energy}. As discussed above, this stems from the blueshift of the outgoing modes of the scalar field near the inner horizon during the regular black hole phase. The perturbed field of the ``long RBH'' model spends more time near the inner horizon; therefore, one expects a greater increase in its final energy. It is important, however, not to identify this blueshift directly with an increase in the field amplitude. As can be seen from Eq.~\eqref{eq:PG_energy}, the scalar-field energy
depends not only on \(\phi\), but also on its radial and time derivatives. This distinction is also familiar from analyses of inner-horizon instabilities, where the scalar field itself can remain finite while its gradients become strongly enhanced~\cite{Ori1997}. This comparison is shown in Fig.~\ref{fig:PG_energy_comparison} for $0<T<100$. Before and after the black-hole phases, the models are in a static phase, which means that Eq.~\eqref{eq:PG_energy} is indeed the conserved energy associated with the scalar field defined on a static PG slicing. The results show energy amplification in both cases, with greater amplification for the model with a longer black-hole phase. 

We have checked the robustness of these results by repeating representative simulations at different radial resolutions. The late-time waveform differences, echo structure, and relative energy amplification remain stable under these changes. Small residual discretization artifacts remain in the regime of strongest blueshift, but they do not affect the qualitative conclusions presented here. We discuss the behavior of the test field near the inner horizon in greater detail below.


\subsection{Inner-horizon blueshift: a closer look}
\label{subsec:inner_horizon_blueshift}

To better understand what happens during the RBH phase, it is useful to follow directly the radial profile of the scalar field. For this purpose, we consider an illustrative BHM--RBH--BHM evolution with the same type of transition as the models discussed above, but with an RBH phase extending approximately over \(T\in[5,50]\). The longer intermediate phase is chosen only to make the evolution near the inner horizon more clearly visible and
should not be identified with either the short or long model considered above.

\begin{figure*}[h]
    \centering

    \includegraphics[width=0.24\textwidth]{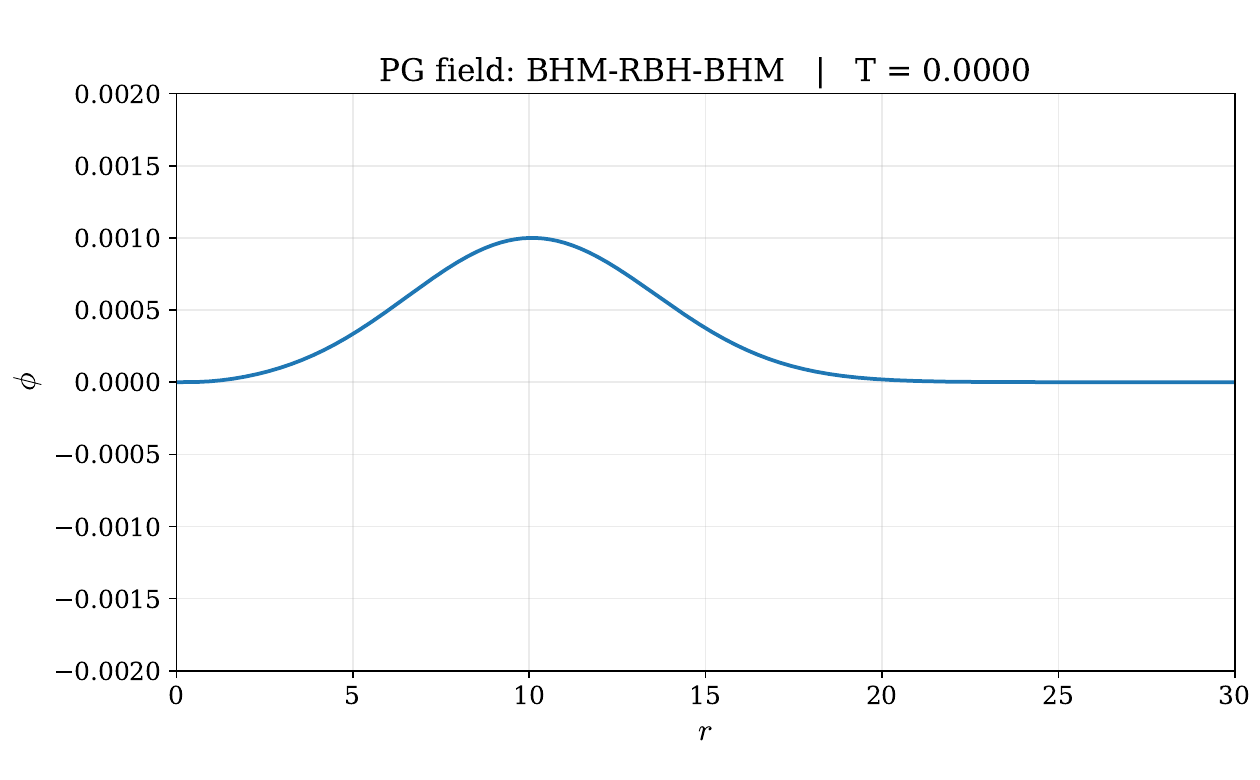}
    \hfill
    \includegraphics[width=0.24\textwidth]{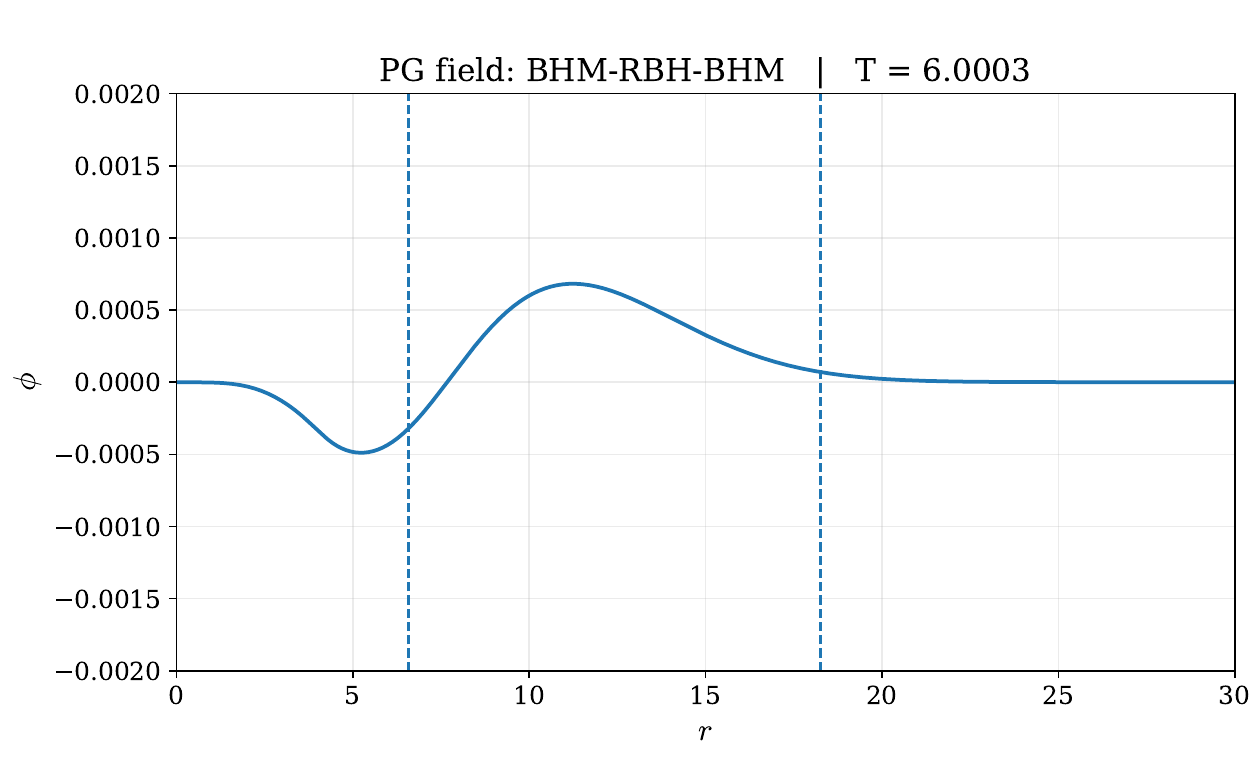}
    \hfill
    \includegraphics[width=0.24\textwidth]{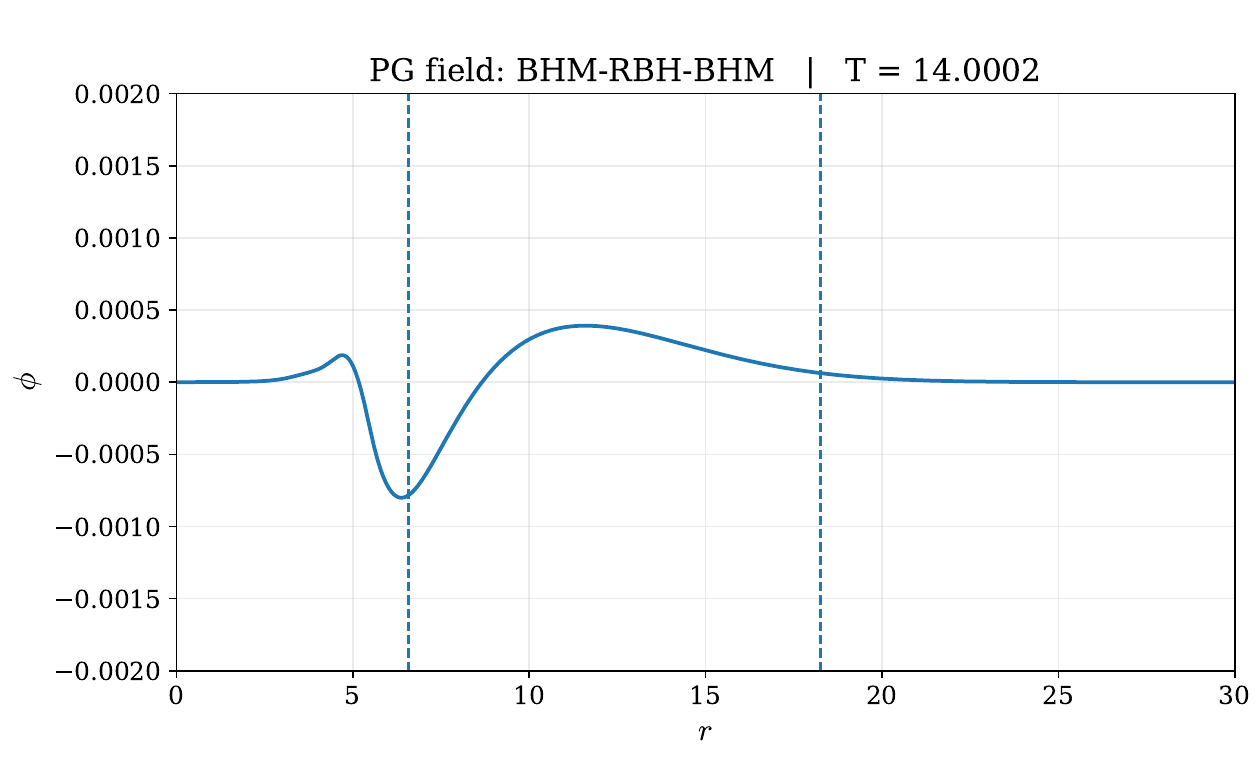}
    \hfill
    \includegraphics[width=0.24\textwidth]{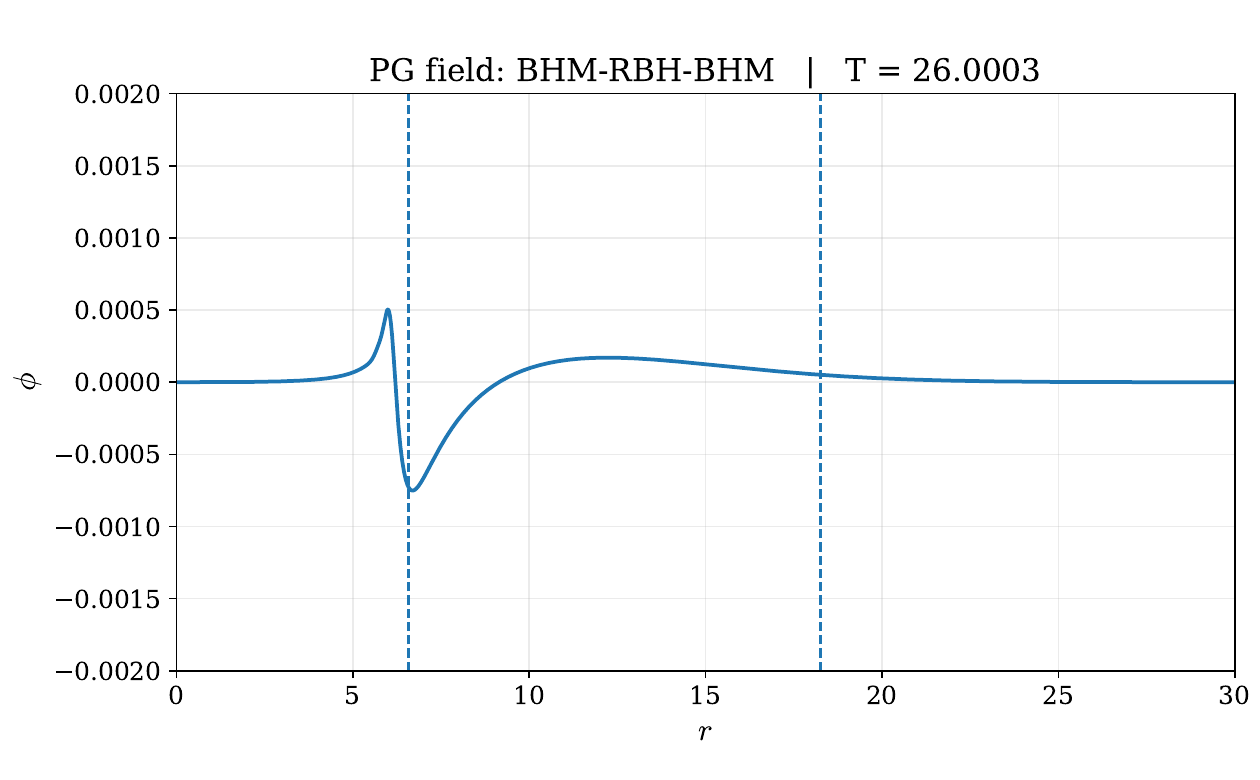}

    \vspace{0.4em}

    \includegraphics[width=0.24\textwidth]{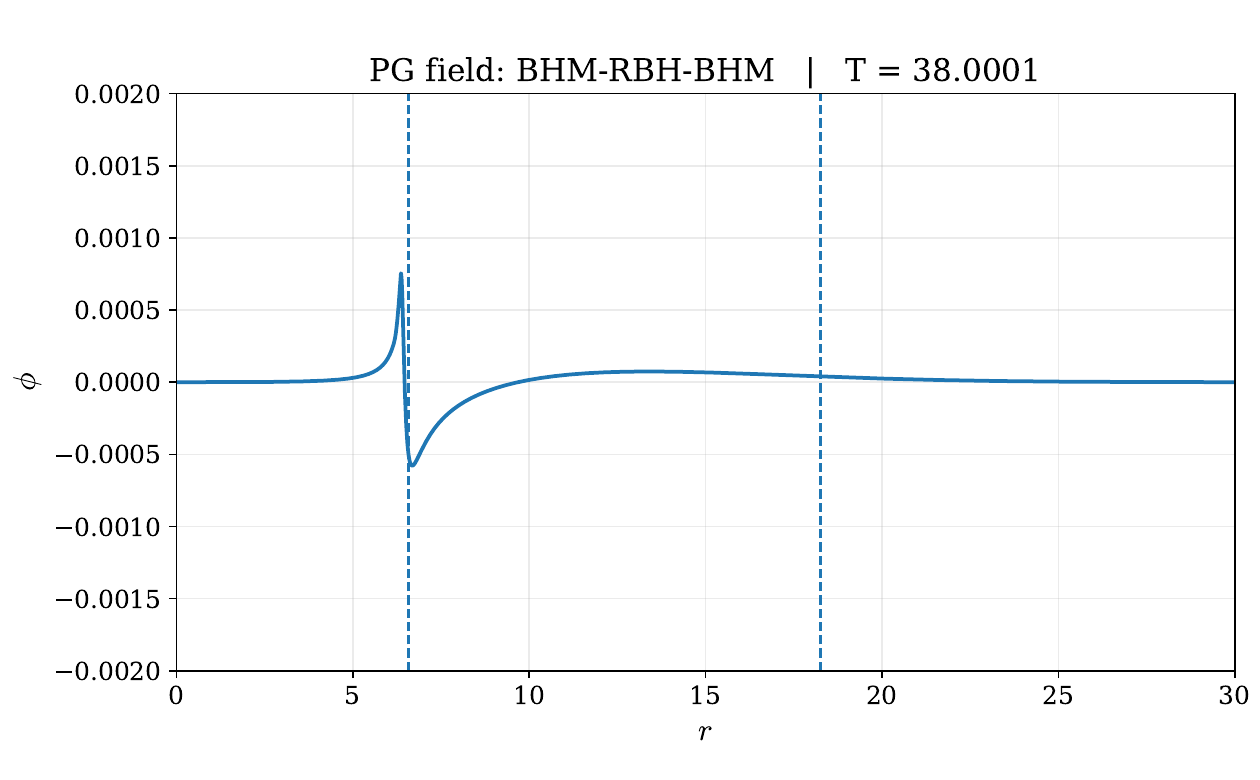}
    \hfill
    \includegraphics[width=0.24\textwidth]{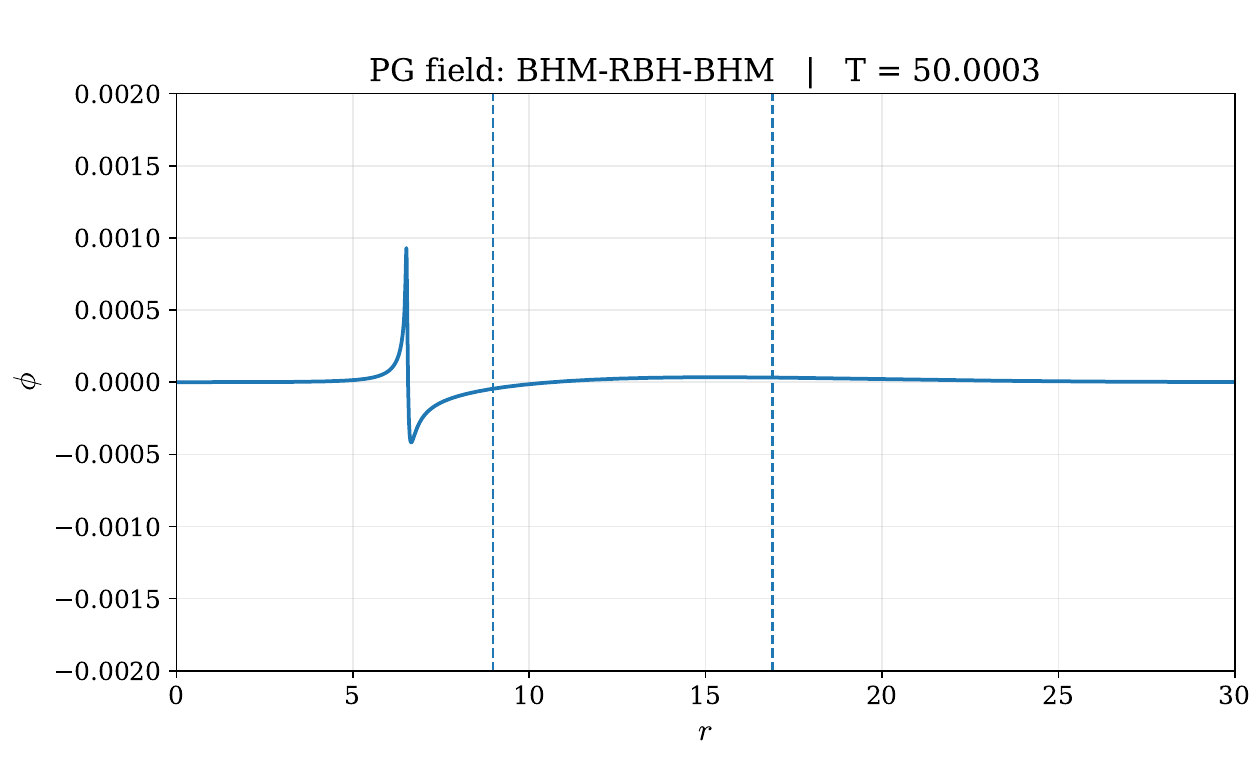}
    \hfill
    \includegraphics[width=0.24\textwidth]{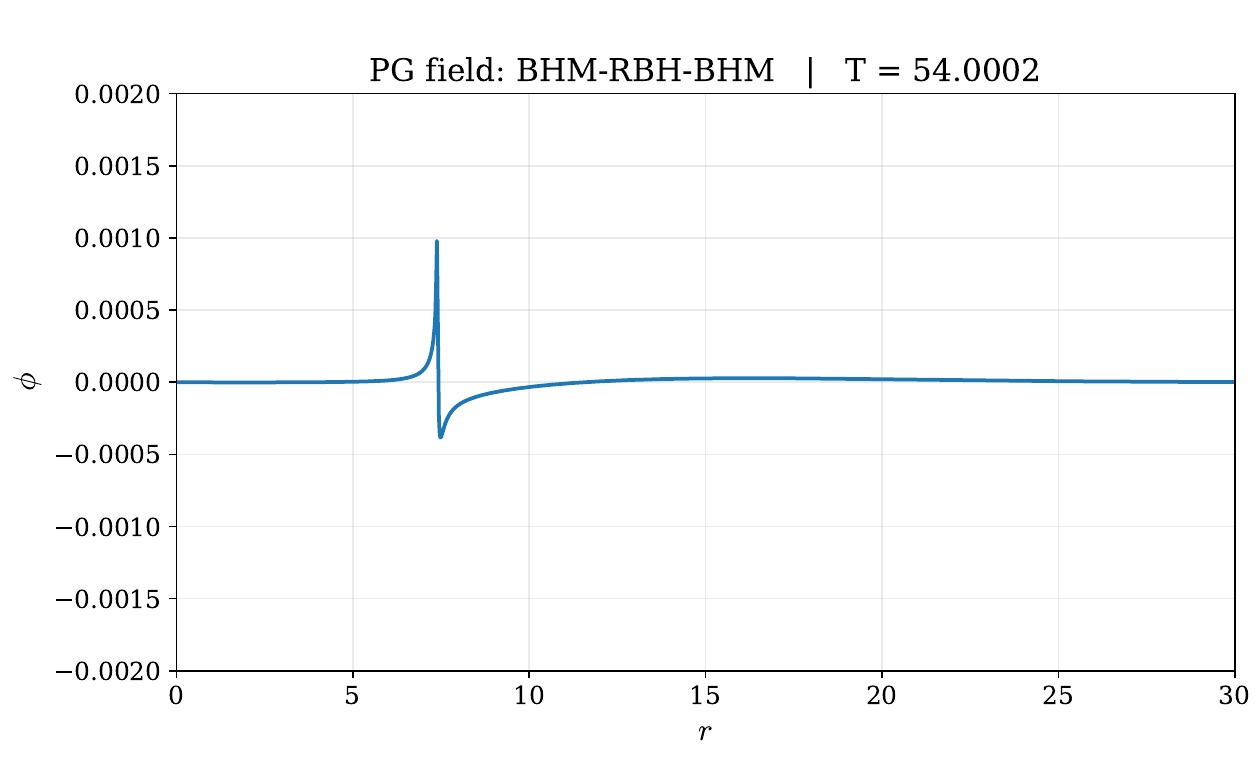}
    \hfill
    \includegraphics[width=0.24\textwidth]{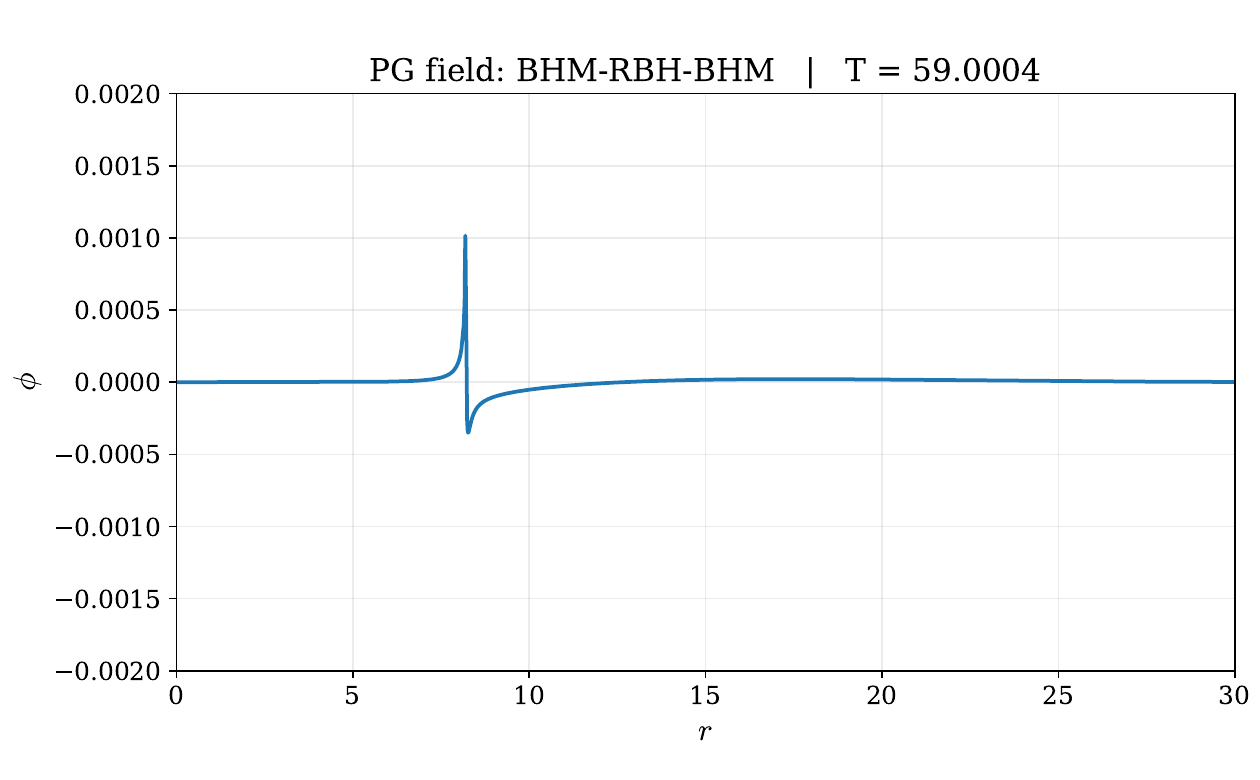}

    \caption{
    Snapshots of the radial scalar-field profile during an illustrative BHM--RBH--BHM evolution. The evolution starts from the initial Gaussian profile. During the RBH phase, the dashed vertical lines indicate the inner and outer horizons. The outgoing part of the perturbation progressively accumulates and sharpens near the inner horizon as a consequence of the blueshift. Once the horizons disappear, the accumulated wave packet is released and begins to propagate outwards.
    }
    \label{fig:PG_inner_horizon_evolution}
\end{figure*}

Figure~\ref{fig:PG_inner_horizon_evolution} shows a sequence of snapshots of the scalar field during this evolution. The system starts from the
Gaussian profile introduced in Eq.~\eqref{eq:initial_gaussian}. As the
geometry enters the RBH phase, an inner and an outer horizon form, indicated by the dashed vertical lines. As the evolution proceeds, the outgoing part
of the wave packet becomes increasingly compressed near the inner horizon, following the tendency of radial null geodesics to approach this surface exponentially fast in $T$-time, which causes an immense blueshift. Such blueshift is visible in the sharpening of the field profile, which changes rapidly in an increasingly narrow region surrounding the inner horizon. 

Towards the end of the RBH phase, the two horizons merge and disappear; the wave packet finds itself in an untrapped region and begins to propagate outwards. At much later times, the released packet would reach the potential barrier of the final BHM configuration, getting scattered there, with part of the radiation transmitted through the barrier and the remaining part reflected back towards the interior. 

Coming back to Fig.~\ref{fig:PG_dynamic_waveform_comparison}, we see that, apart from the very small early-time oscillations associated with the initial data, this released burst is the first significant signal to cross the potential barrier outwards and corresponds mainly to the first peak in the observed field amplitude. The inner-horizon blueshift then transfers spectral power towards higher frequencies. These higher-frequency components are transmitted more efficiently through the BHM potential barrier, resulting in a larger first peak. At the same time, a smaller fraction of the wave packet is reflected back into the interior, leaving less energy available to generate subsequent echoes.  We have thus identified a waveform feature that is directly associated with what takes place inside a transient trapped region.


\subsection{Oscillating BHM--RBH--BHM}
\label{subsec:Oscillating}

Lastly, we consider the oscillating BHM--RBH--BHM scenario introduced in Sec.~\ref{sec:Models}, in which the geometry crosses the threshold of horizon formation several times before relaxing to a static configuration. As an example, here we consider BHMs undergoing three temporary and equally-spaced RBH phases before returning to their initial configuration. In contrast with the comparison of the previous subsection, where we varied the duration of a single RBH phase, here we keep the duration of each RBH phase fixed and vary the amount of time spent in the intermediate BHM configuration between successive RBH phases.

Both models have \(\ell=2\), \(M=10\),
\(L_\textsc{bhm}=1.03\,L_{\rm crit}\), and
\(L_\textsc{rbh}=0.70\,L_{\rm crit}\). Each temporary RBH phase lasts
approximately five time units. In the ``short-separation'' model, the BHM interval between successive RBH phases is \(5\), giving RBH phases approximately over
\[
    T\in[5,10]\cup[15,20]\cup[25,30].
\]
In the ``long-separation'' model, this interval is instead \(300\), so that the three RBH phases occur approximately over
\[
    T\in[5,10]\cup[310,315]\cup[615,620].
\]
Apart from this difference in the separation between successive RBH phases, the numerical setup and initial data are the same in the two
cases.

\begin{figure}[h]
    \centering
    \includegraphics[width=0.70\linewidth]
    {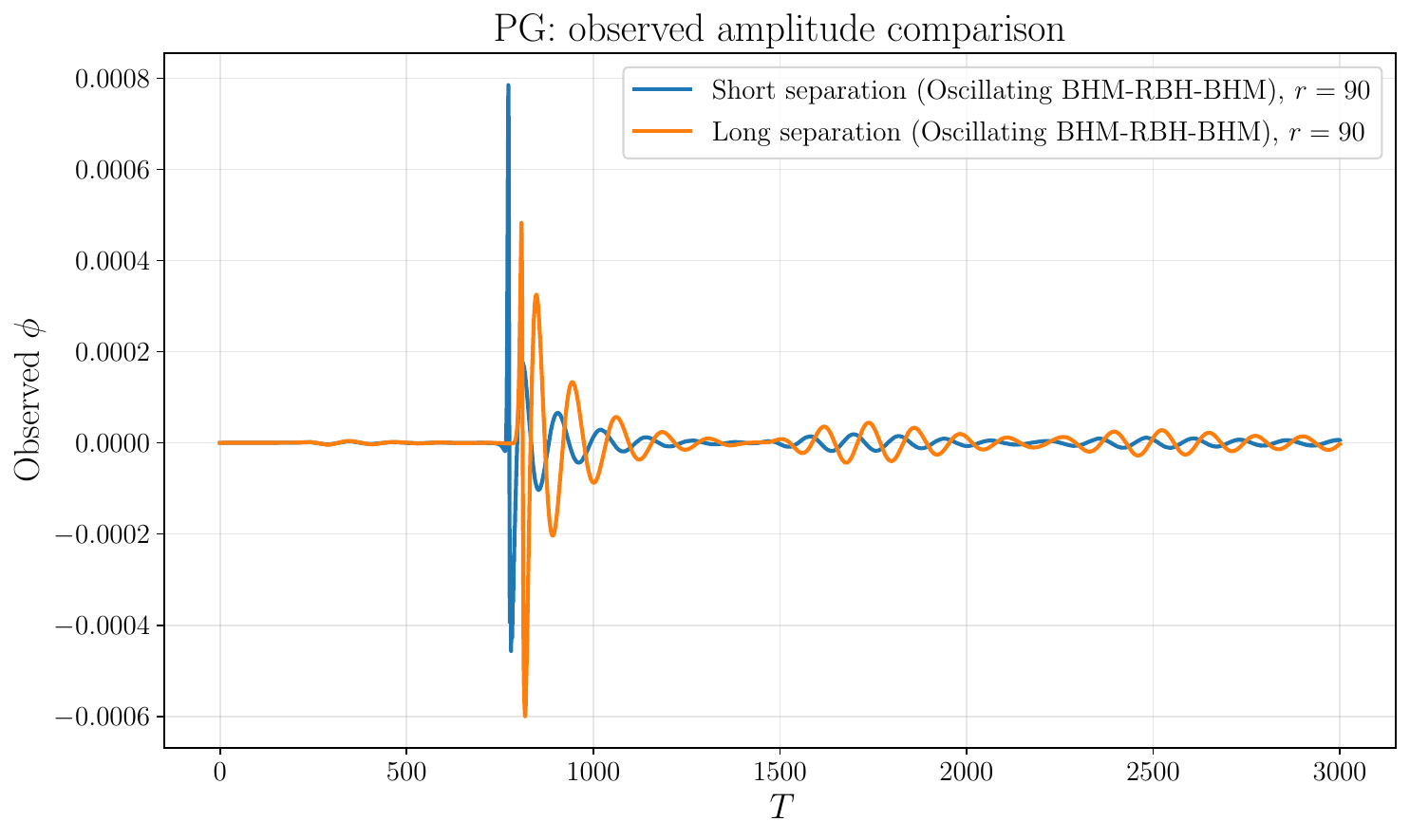}
    \caption{
    Observed scalar amplitude in PG coordinates for two oscillating
    BHM--RBH--BHM evolutions measured at \(r_{\rm obs}=90\).
    Both models have \(\ell=2\), \(M=10\),
    \(L_\textsc{bhm}=1.03\,L_{\rm crit}\), and
    \(L_\textsc{rbh}=0.70\,L_{\rm crit}\), and undergo three temporary RBH
    phases of the same duration. The two cases differ only in the duration
    of the intermediate BHM phases separating successive RBH episodes:
    \(5\) time units for the short-separation case and \(300\) time units
    for the long-separation case.
    }
    \label{fig:PG_oscillating_separation_comparison}
\end{figure}

Figure~\ref{fig:PG_oscillating_separation_comparison} shows the scalar waveform observed at \(r_{\rm obs}=90\). The two signals differ substantially even though they undergo the same number of RBH phases with the same individual duration. In particular, the dominant first burst in the long-separation case reaches the observer later than in the short-separation case. In the latter, all three RBH phases are completed very early in the evolution, after which the perturbation propagates on the final BHM background. For the scalar waveform considered here, the rapid succession of the three RBH episodes produces a response qualitatively similar to that of a single long-lived trapped region, since the individual transitions are not clearly
resolved in the distant signal. In the long-separation case, instead, the second and third RBH phases occur while the perturbation is still evolving in the compact region. The formation and disappearance of these later trapped regions therefore modify the subsequent propagation of the field and delay the release of the dominant outgoing signal.

The comparison shows that the waveform is sensitive not only to the duration and number of temporary trapped regions, but also to their timing relative to the evolution of the perturbation. Changing the separation between otherwise identical RBH phases changes both the arrival time and the amplitude of the main outgoing burst. In principle, there could be situations in which long-lived perturbations that remain trapped within a BHM entering a transient RBH phase get amplified and then released, leading to an increase in the amplitude of late echo signals. The oscillating histories therefore provide another example in which the final signal retains information about the detailed dynamical history of the interior. It should be noted that, depending on how close to extremality the final BHM is, the field might remain nearly frozen around the would-be horizon, thus taking an enormous amount of time (as seen by a faraway observer) to escape to infinity. This carries implications for the feasibility of detecting gravitational-wave echoes as part of early ringdown signals~\cite{Marketal2017} or as independent, isolated events~\cite{Zimmermanetal2023}.


\section{Conclusions}

The classical and semiclassical instabilities of inner horizons call into question the description of black-hole interiors as stationary configurations. In a nonsingular setting, these instabilities may instead drive a dynamical evolution involving the formation and disappearance of trapped regions. The phenomenological consequences of this evolution need not remain hidden from distant observers. When trapping horizons are temporary, radiation that has interacted with the interior can escape, carrying information about the evolution of the geometry. This enables the possibility of extracting information from strong-gravity regimes inaccessible in stationary black-hole spacetimes and, if indeed semiclassical instabilities work in the direction of depleting the trapped region, it might even open a new channel for testing quantum field theory in curved spacetimes in astrophysical scenarios.

In this work, we have investigated the phenomenology of transient trapped regions by evolving linear scalar perturbations on prescribed, time-dependent Hayward-like backgrounds. We considered BHM--RBH--BHM histories in which an initially horizonless ultra-compact object develops a compact trapped region before returning to a horizonless configuration, together with histories containing repeated transitions of this kind. These constructions provide phenomenological models of the relaxation processes suggested by recent studies of semiclassical backreaction. Building on the analysis of Cardoso et al.~\cite{Cardoso:2023guh}, we followed both the scalar-field energy and the time-dependent waveform. The use of horizon-penetrating Painlev\'e--Gullstrand coordinates allowed us to prescribe Cauchy initial data and evolve the perturbations through the formation and disappearance of trapping horizons within a single framework.

Our main finding is that the late-time signal retains a memory of the intermediate trapped-region phase. Evolutions with identical initial and final geometries, differing only in the duration of this phase, produce waveforms with different amplitudes and phases. In the cases studied, a longer trapped phase also leads to greater energy amplification, consistent with the blueshift of outgoing modes near the inner trapping horizon. The horizonless configurations support echoes arising from repeated scattering between the regular interior and the outer potential barrier, while the dynamical transitions modify the resulting late-time response. Repeated BHM--RBH--BHM transitions likewise modify the waveform, with the timing of successive trapped phases leaving a distinct imprint on the
outgoing signal. More generally, the energy amplification and the waveform structure provide complementary information about the evolution of the compact object.

Since our models are formulated in geometric units, the numerical value of the ADM mass does not correspond to a fixed physical mass scale. As an illustrative example, if the value $M=10$ used in our simulations is identified with a physical mass of $30M_\odot$, representative of the stellar-mass black holes observed by LIGO--Virgo--KAGRA, the short and long RBH phases considered here, with durations $T_{\rm w}=5$ and
$T_{\rm w}=15$, correspond to approximately
$7.4\times10^{-5}\,\mathrm{s}$ and
$2.2\times10^{-4}\,\mathrm{s}$, respectively. Although these intervals are very short on astrophysical scales, the instability associated with a non-extremal inner horizon grows exponentially with time. Consequently, even a short-lived RBH phase can, in principle, generate sufficiently strong amplification to induce non-negligible dynamics in the interior. The shortness of the RBH phase therefore does not by itself justify treating the interior as stationary, and this mechanism can remain relevant for astrophysical black holes.  A possible alternative that could halt this evolution is relaxation towards an inner-extremal configuration, whose vanishing inner-horizon surface gravity suppresses the exponential growth mechanism~\cite{Carballo-Rubioetal2022c, Carballo-Rubio:2026gwg}. Such configurations, and their possible role as endpoints of the evolution, lie beyond the scope of the present analysis. 

Our results should be understood within the phenomenological and test-field approximations adopted here. The background histories are prescribed, and the scalar perturbations do not backreact on the geometry. We therefore establish how temporary trapped regions can leave an imprint on outgoing radiation without determining self-consistently which histories are realised or predicting their gravitational-wave signatures. Extending the analysis to rotating configurations, gravitational perturbations, and a coupled evolution of the geometry and its perturbations will be necessary to assess the observational relevance of this mechanism. Similarly, this analysis motivates the need to understand the evolution of black hole interiors in general relativity and semiclassical gravity better, with the aim of improving gravitational collapse models and determining whether the observational features identified in this work are retained in more realistic scenarios. 

These findings motivate a broader approach to black-hole phenomenology in which the relaxation history of the compact object is included alongside the properties of its final configuration. In particular, interpreting late-time radiation using stationary backgrounds alone may overlook information carried by the amplitudes, phases, and echo structure of the signal. We therefore hope that the present investigation will stimulate further studies connecting the dynamics of black-hole interiors to observable radiation through the full evolution of the spacetime.


\begin{acknowledgments}
The authors wish to thank Jacopo Mazza for useful discussions during the development of this work. 
\end{acknowledgments}


\appendix


\section{Scalar evolution in Eddington--Finkelstein coordinates}
\label{app:EF_scalar}

For completeness, we briefly summarize the ingoing
Eddington--Finkelstein formulation used in the static comparison of
Sec.~\ref{subsec:static_uco}. For a static spherically symmetric
background, the metric can be written as
\begin{equation}
    ds^2
    =
    -F(r)\,dv^2
    +
    2\,dv\,dr
    +
    r^2d\Omega^2 ,
\end{equation}
where \(v\) is the ingoing null coordinate. Using the decomposition
\begin{equation}
    \Phi(v,r,\theta,\varphi)
    =
    \frac{\phi(v,r)}{r}
    Y_{\ell m}(\theta,\varphi),
\end{equation}
the massless scalar equation becomes
\begin{equation}
    2\,\partial_v\partial_r\phi
    +
    \partial_r\!\left(F\,\partial_r\phi\right)
    -
    \left[
        \frac{F'}{r}
        +
        \frac{\ell(\ell+1)}{r^2}
    \right]\phi
    =
    0 .
    \label{eq:EF_scalar_appendix}
\end{equation}

For the numerical evolution, it is convenient to introduce
\begin{equation}
    \psi(v,r)=\partial_r\phi(v,r),
\end{equation}
so that Eq.~\eqref{eq:EF_scalar_appendix} can be written as
\begin{equation}
    \partial_v\psi
    =
    -\frac{1}{2}
    \partial_r(F\psi)
    +
    \frac{1}{2}
    \left[
        \frac{F'}{r}
        +
        \frac{\ell(\ell+1)}{r^2}
    \right]\phi .
\end{equation}

The main difference with the PG formulation concerns the initial data. A surface of constant \(v\) is null rather than spacelike, so one cannot independently prescribe both the field and its time derivative. We therefore specify the same regularized Gaussian radial profile used in the PG simulations on the initial surface \(v=0\), while
\(\psi(0,r)\) is fixed by its radial derivative. This corresponds to outgoing initial data. Regularity at the center is imposed in the same way as in the PG evolution. The EF calculation therefore provides a useful independent check of the echo structure, although its initial-value problem is not identical to the Cauchy problem used in PG coordinates.


\section{Scalar evolution in tortoise coordinates}
\label{app:tortoise_scalar}

For the static BHM considered in Sec.~\ref{subsec:static_uco}, the scalar
field can also be evolved using the usual static time coordinate \(t\) and the tortoise coordinate \(r_*\), defined by
\begin{equation}
    \frac{dr_*}{dr}
    =
    \frac{1}{F(r)} .
\end{equation}
Using the same scalar decomposition introduced in
Eq.~\eqref{eq:scalar_decomposition}, the Klein--Gordon equation takes the
standard form
\begin{equation}
    \partial_t^2\phi
    -
    \partial_{r_*}^2\phi
    +
    V_\ell(r)\phi
    =
    0,
    \label{eq:tortoise_scalar_appendix}
\end{equation}
where the effective potential \(V_\ell(r)\) is given in
Eq.~\eqref{eq:scalar_effective_potential_models}.

This formulation makes the scattering interpretation particularly transparent: the scalar pulse propagates in the effective potential \(V_\ell\), and the echo signal results from repeated propagation between the regular interior and the outer potential barrier.

For the comparison shown in Sec.~\ref{subsec:static_uco}, the initial radial profile is the same regularized Gaussian introduced in Eq.~\eqref{eq:initial_gaussian}, expressed as a function of \(r_*\) through \(r=r(r_*)\). We choose time-symmetric initial data,
\begin{equation}
    \partial_t\phi(0,r_*)=0 .
\end{equation}
Since this formulation is used only for the static horizonless configuration, no horizon is encountered in the numerical domain.


\bibliographystyle{apsrev4-1}
	\bibliography{bib-refs}


\end{document}